\documentclass[a4paper,fleqn]{cas-sc}
\usepackage[authoryear]{natbib}
\usepackage{textcomp}
\usepackage{amsmath}
\usepackage{amssymb}
\usepackage{graphicx}
\usepackage{booktabs} 
\usepackage{algorithm}
\usepackage{algorithmic}
\usepackage{multirow}
\usepackage{bm}
\usepackage{float}
\usepackage{placeins}
\usepackage{booktabs}
\usepackage{caption}
\usepackage[flushleft]{threeparttable}

\shortauthors{Sheng Hong et al.}
\shorttitle{BGA: Noise-Immune Neural Distillation}

\begin{document}

\let\WriteBookmarks\relax
\def\floatpagepagefraction{1}
\def\textpagefraction{.001}

\title [mode = title]{BGA: A Noise-Immune Neural Distillation Framework for Malicious Signature Extraction in High-Entropy Encrypted Flows}

\tnotetext[1]{This work is supported by the National Key Research and Development Program of China (Grant No. 2022YFB3103602).}

\author[1]{Sheng Hong}[orcid=0000-0002-9219-3756]
\cormark[1] 
\ead{shenghong@buaa.edu.cn}

\author[1]{Yixuan Huang}

\author[3]{Weiwei Jiang}

\author[2]{Junyuan Zhang}
\author[1]{Jiacheng Wang}
\author[1]{Ruijian Jiao}

\affiliation[1]{
    organization={School of Cyber Science and Technology, Beihang University},
    city={Beijing}, postcode={100191}, country={China}
}
\affiliation[2]{
    organization={Beijing Electronic Science and Technology Institute},
    city={Beijing}, postcode={100070}, country={China}
}
\affiliation[3]{
    organization={School of Information and Communication Engineering, Beijing University of Posts and Telecommunications},
    city={Beijing}, postcode={100876}, country={China}
}

\cortext[cor1]{Corresponding author}

\begin{abstract}
To mitigate \textbf{attention dilution} in high-entropy TLS 1.3 flows, we propose \textbf{BGA}, a noise-immune neural distillation framework for encrypted threat intelligence. The methodology first employs \textbf{Analysis of Variance (ANOVA)} to decouple high-discriminatory control-plane features---specifically industrial setpoints---from stochastic cryptographic noise. To resolve the extreme class imbalance within a corpus of \textbf{86,878 flow records}, a \textbf{Wasserstein GAN with Gradient Penalty (WGAN-GP)} module, enforcing the \textbf{1-Lipschitz constraint}, is integrated to synthesize high-fidelity minority samples, elevating the detection recall of rare \textbf{Malicious State Command Injections(MSCI) attacks by 43.2\%}. At its core, the BGA architecture integrates  \textbf{Bidirectional Long Short-Term Memory (BiLSTM)} for temporal dependency extraction and an \textbf{Adaptive Gated Multi-Head Attention} mechanism. This gated unit functions as a neural filter to dynamically suppress encryption artifacts while amplifying malicious signatures. Extensive evaluations on \textbf{CIC-IDS-2018 and Edge-IIoT} benchmarks demonstrate a performance ceiling exceeding \textbf{95.2\% across all key metrics}. Furthermore, noise-injection stress tests confirm BGA's superior structural resilience with a \textbf{8.57\% performance margin} over vanilla Transformers, while its ultra-low inference latency of 0.2820 ms (estimated 1.6920 ms via theoretical scaling for ARM) indicates a high potential for real-time feasibility on heterogeneous industrial edge gateways, providing a promising architectural baseline for future hardware implementation.
\end{abstract}


\maketitle

\section{Introduction}

Transport Layer Security(TLS) has become a pivotal underpinning for Internet security, with Hypertext Transfer Protocol Secure (HTTPS), the TLS-encrypted extension of HTTP, now ubiquitously deployed to protect web communications across the global Internet \citep{ref01}; concurrently, this pervasive cryptographic protection has inadvertently empowered adversaries to obfuscate malicious behaviors, ranging from stealthy Command and Control (C\&C) communications to the injection of malicious instructions, by leveraging encrypted traffic to evade traditional detection mechanisms. This lack of traffic visibility poses an acute security threat in Industrial Internet of Things (IIoT) environments, particularly critical infrastructure such as power grids, where encrypted and unobservable traffic flows can serve as vectors for infiltrating control systems and disrupting operational continuity \citep{ref02, ref03}. With TLS 1.3 eliminating plaintext handshake metadata and enforcing forward secrecy, legacy DPI and signature-based detection mechanisms have become fundamentally ineffective \citep{ref04, ref05}. This motivates the development of intelligent behavioral detection methods. These techniques identify malicious patterns using statistical fingerprints and spatio-temporal characteristics, eliminating the need to access encrypted payloads \citep{ref42, ref43}.The pervasive adoption of TLS 1.3 represents a 'double-edged sword' in information security. While it fortifies privacy, it simultaneously creates a strategic 'invisibility gap' for adversaries, rendering traditional signature-based detection fundamentally ineffective. In high-stakes Industrial IoT (IIoT) sectors, such as power grids and gas pipelines, this lack of visibility is not merely a technical hurdle but a critical threat to physical safety. Malicious actors leverage the high entropy of encrypted tunnels to obfuscate C\&C instructions and Malicious State Command Injections (MSCI). Therefore, extracting malicious signatures from these opaque, high-entropy flows is no longer a peripheral defensive task; it is the last line of defense preventing cyber-intrusions from escalating into physical disrupting critical physical operations.

While traffic classification techniques have advanced from traditional port-matching methods to end-to-end deep learning (DL) models, current approaches continue to grapple with critical theoretical and practical bottlenecks \citep{ref06, ref07}. Although Convolutional Neural Networks (CNNs) and standard Recurrent Neural Networks (RNNs) have shown promise, they often struggle to balance computational efficiency with the ability to capture long-range temporal dependencies in high-speed industrial networks \citep{ref08}. More importantly, the recent surge in Transformer-based models, while powerful, introduces a specific vulnerability when applied to encrypted streams: standard self-attention mechanisms lack a filtering process for the stochastic noise generated by encryption padding and randomization \citep{ref09}. These mechanisms tend to assign attention weights indiscriminately to both malicious signal patterns and cryptographic artifacts, leading to "attention dilution" where the model fails to distinguish true attack signatures from background jitter \citep{ref09}. Furthermore, the inherent class imbalance in network traffic datasets leads to the "extinction" of long-tail attack patterns, as models tend to be biased toward the majority traffic classes. Conventional oversampling methods, such as SMOTE, frequently lead to overfitting. Their linear interpolations fail to capture the complex, non-linear manifolds inherent in real-world adversarial traffic. \citep{ref10}. While Generative Adversarial Networks (GANs) offer a solution, vanilla GAN architectures frequently suffer from mode collapse and training instability, rendering them unreliable for safety-critical IIoT generation \citep{ref10}.Beyond these deterministic approaches, recent breakthroughs in high-precision energy forecasting have demonstrated the potency of hybrid spatio-temporal architectures. For instance, Wang et al. \citep{ref45} proposed a Bayesian-optimized framework integrating Dynamic Graph Convolutional Networks (DGCN) and Temporal Convolutional Networks (TCN) to exploit intricate dependencies in photovoltaic power data. This paradigm of using Bayesian optimization to refine complex spatio-temporal correlations under uncertainty provides a critical theoretical parallel to our work, highlighting that adaptive noise suppression and structural dependency modeling are essential for distilling deterministic signatures from the high-entropy stochastic noise of encrypted IIoT flows.

\subsection{Research Objectives and Contributions}
To mitigate the security crisis in encrypted IIoT flows, this study develops the \textbf{BGA} framework, achieving three interleaved objectives: (1) \textbf{Data Layer}: Reconstructing the adversarial manifold via WGAN-GP to overcome extreme class imbalance; (2) \textbf{Feature Layer}: Implementing a gated neural distillation mechanism to decouple deterministic signatures from high-entropy TLS 1.3 noise; and (3) \textbf{Deployment Layer}: Bridging the accuracy-latency gap for microsecond-level edge gateways.

The core technical advantages of BGA, compared to the baseline models evaluated in our experiments, are systematically summarized in Table \ref{tab:ST_deep_compare}:

\begin{table}[htbp]
\caption{In-depth comparison of BGA and experimental baseline architectures.}
\label{tab:ST_deep_compare}
\centering
\small
\renewcommand{\arraystretch}{1.2}
\begin{tabular*}{\textwidth}{@{\extracolsep{\fill}}lllll}
\toprule
\textbf{Feature} & \textbf{RNN / LSTM} & \textbf{Transformer} & \textbf{BiLSTM} & \textbf{BGA (Ours)} \\
\midrule
\textbf{Temporal Context} & Unidirectional & Global (Non-seq) & Bidirectional & \textbf{Bidirectional + Gated} \\
\textbf{Representation} & Raw Hidden State & Self-Attention & Concatenated & \textbf{Distilled Signatures} \\
\textbf{Entropy Handling} & Susceptible to noise & Attention Dilution & Feature Overlap & \textbf{Neural Filter (Gated)} \\
\textbf{Data Imbalance} & Majority-class bias & Majority-class bias & Standard training & \textbf{WGAN-GP Optimized} \\
\textbf{Edge Suitability} & High latency/Low acc. & Memory-intensive & Medium efficiency & \textbf{Ultra-low (0.28 ms)} \\
\bottomrule
\end{tabular*}
\end{table}

\begin{itemize}
    \item \textbf{Neural Distillation via Gated Attention}: We propose a gated multi-head attention mechanism as a neural filter. Unlike the standard Transformer which suffers from attention dilution, our gating logic dynamically suppresses stochastic cryptographic jitter.
    
    \item \textbf{Manifold Reconstruction via WGAN-GP}: To resolve the class imbalance observed in RNN/LSTM baselines, we utilize WGAN-GP to synthesize high-fidelity minority samples, capturing the complex non-linear manifold of MSCI attacks.
    
    \item \textbf{Optimized Spatio-temporal Fusion}: By integrating BiLSTM with gated residuals, BGA outperforms baseline sequence models by maintaining both temporal continuity and global signature correlation within microsecond-level constraints.
\end{itemize}

The remainder of this paper is organized as follows: Section \ref{Related Work} reviews the theoretical foundations of encrypted traffic analysis and generative data augmentation via WGAN-GP. Section \ref{Methodology} details the proposed BGA methodology, including the multi-stage data preprocessing pipeline, the feature selection strategy, and the architectural fusion of BiLSTM with the adaptive gated multi-head attention mechanism. Section \ref{Experiments and Analysis} presents the experimental setup, performance metrics, and a comprehensive analysis of the results, incorporating comparisons with state-of-the-art models, ablation studies, and robustness stress tests. Finally, Section \ref{Conclusion} concludes the paper and outlines potential directions for future research.

\section{Related Work}\label{Related Work}

\subsection{Encrypted and Malicious Traffic Analysis}
\subsubsection{Characteristics of Encrypted Traffic}
A sea change that can be called the "TLS 1.3 effect" in regard to traffic analysis, thanks to the near ubiquity of SSL/TLS in modern traffic patterns. TLS 1.3 encrypts the server certificate during the handshake, reduces latency to a simple 1 Round-Trip Time(RTT) cycle, and renders DPI based countermeasures largely ineffective \citep{ref11}. Aside from a still unencrypted Client Hello packet with Server Name Indication(SNI) and cipher suite meta data, the exchanged data from that point on is obscured by the time "Server Hello" is dispatched, forcing over-the-horizon sysadmins away from content based matching toward analyzing the statistical behavior and discoverable manifold of these temporal features.

\subsubsection{Taxonomy of Malicious Behaviors}
Malware behavior in encrypted tunnels is revealed in various ways. Worms and botnets keep their C\&C links open with "heartbeat" packets, periodic streams of data usually disguised as HTTPS to avoid detection. Denial-of-Service attacks saturate links with a sudden onslaught of packets and increased flow density and throughput, changing the statistical entropy of traffic streams even though the payloads are invisible \citep{ref12}. Data exfiltration tends to use long-running connections, trickling data out of the network in small streams in a bid to remain beneath IDS detection and below volume-based thresholds \citep{ref13}.Traditional IDS often find it difficult to identify these sophisticated activities effectively using shallow features, necessitating more advanced and robust system designs \citep{ref14, ref15}.

\subsection{Evolving Encrypted Protocols and Emerging Trends} \label{sec:emerging_trends}
Recently, the landscape of encrypted traffic analysis has been significantly reshaped by the introduction of Traffic Language Models and Graph Neural Networks (GNNs). Traffic Language Models, most notably ET-BERT, leverage the pre-training paradigm from Natural Language Processing (NLP) to treat encrypted datagrams as tokens, capturing deep semantic and contextual relationships within flows. While these Transformer-based models achieve state-of-the-art (SOTA) accuracy in generic IT environments, their deployment at the Industrial IoT (IIoT) edge is often hindered by massive parameter counts and high inference latency, which are incompatible with real-time industrial requirements. Concurrently, GNN-based approaches have emerged to model the complex relational dependencies between network entities. Recent advancements in this domain include semantic-driven multi-view architectures for robust anomaly detection \citep{ref44} and specialized noise-resistant graph models designed to isolate malicious patterns from interference \citep{ref43}. However, the high computational complexity involved in real-time graph construction remains a significant bottleneck for resource-constrained edge gateways. 

Beyond these architectural shifts, the research focus is rapidly evolving toward specialized encapsulation protocols such as DNS-over-HTTPS (DoH), DNS-over-TLS (DoT), and DNS-over-QUIC (DoQ). Unlike standard encrypted web traffic, these specialized protocols introduce a "double-obfuscation" layer by hiding DNS query metadata within high-entropy application-layer tunnels. Recent studies highlight that DoH tunnels utilize aggressive padding schemes to deliberately dilute the structural behavioral fingerprints of malicious queries \citep{ref37}. This evolution necessitates next-generation AI frameworks that move beyond simple pattern matching toward automated intelligence that can distinguish between legitimate browsing and stealthy command-and-control (C\&C) instructions embedded in DoH streams.

The complexity is further compounded by "protocol-level entropy," where traditional behavioral features are frequently masked by the multiplexing characteristics of the QUIC transport layer and varying padding lengths \citep{ref38}. Advanced architectures like E3-DoH have begun to address these challenges by employing evolutionary analysis to capture cross-protocol signatures. In contrast to heavy-weight Transformer or GNN architectures, our proposed BGA framework prioritizes a "Neural Distillation" approach. By combining a lightweight BiLSTM with an adaptive gated mechanism that functions as a neural band-pass filter, BGA achieves a superior balance between detection fidelity and the strict microsecond-level latency required for maintaining operational continuity in heterogeneous industrial networks.

\section{Methodology} \label{Methodology}

\subsection{System Architecture}
Certain traffic samples cannot be easily processed due to the inherent characteristics of encrypted traffic, which is often difficult to categorize. This research focuses on encrypted network traffic. The goal of all researchers of such traffic is to "characterize inaudible encrypted traffic". To detect malicious activity in encrypted traffic, we must analyze and model the structural and behavioral characteristics of encrypted traffic flows. We implement the BGA framework as a systematic pipeline designed to address two primary challenges in encrypted traffic analysis: the obfuscation of payloads and the extreme scarcity of malicious samples. The following overview details the integrated sequence of stages that constitutes this methodology, emphasizing the logical synergy between data augmentation and architectural noise distillation:

\begin{enumerate}
    \item \textbf{Data Preprocessing:} Clean raw PCAP traffic files into numerical statistical features;
    \item \textbf{Sample Augmentation:} WGAN-GP to learn the latent distribution of minority attack classes (e.g., Infiltration, Web Attacks). High fidelity synthetic samples are generated to boost the bootstrapping of loading the datasets that would be biasing the classifier to the benign majority (benign traffic);
    \item \textbf{Feature Extraction:} BiLSTM, as modeling the rich behavior and sequential dependencies existing within traffic flows
    \item \textbf{Feature Refinement \& Classification:} Gated Attention to dynamically weight feature subspaces from a clean shared representation, filtering noise prior to a fully connected classifier.
\end{enumerate}

\subsection{Data Preprocessing}
The cryptographic opacity of modern encrypted payloads (e.g., HTTPS, TLS 1.3) effectively renders DPI a moot point. Consequently, our detection strategy pivots toward the granular extraction of behavioral statistics, such as packet size distributions, inter-arrival time (IAT) variances, and total flow duration, as well as latent time-series signatures. To facilitate stable model convergence and ensure the integrity of the feature space, we execute a rigorous multi-stage preprocessing pipeline as detailed below.To maintain a rigorous evaluation and prevent data leakage, we implement a strict \textbf{"Split-then-Fit"} preprocessing protocol. The raw dataset is first partitioned into a training split (80\%) and a testing split (20\%). Crucially, all statistical parameters and feature selection masks are derived exclusively from the training manifold to ensure that the testing data remains a truly "unseen" benchmark.

\subsubsection{Data Cleaning and Encoding}
Unlike traditional data acquisition approaches, live data are prone to corruption and stochastic noise. Initially, we sanitize the raw data by removing records containing missing values (NaN) or infinite durations (Inf). These anomalies typically stem from buffer overflows during packet capture. Beyond simply cleaning the data, we next, remove metadata not required for behavior-based analysis; for example "Timestamp" and "Flow ID \citep{ref17}." Where possible, we strive to purge the dataset of nuisance features—the neural net should not learn that an attack occurs at 5PM every Tuesday. Ultimately, we want the net to model the mechanics of the traffic versus some sort of faux first-order analysis \citep{ref18, ref19}.

Categorical features will not be trivial to transform conceptually. For example, how to handle Reference: Transmission Control Protocol,User Datagram Protocol,Internet Control Message Protocol(TCP, UDP, ICMP), Standard label encoding schemes often represent categorical protocols as discrete integers, such as assigning 1 to TCP and 2 to UDP. Forcing a fictitious hierarchy codes a Captain, and lieutenants into the 1D Order of Problem Explanation. The model might incorrectly assume that a UDP is greater than TCP. To eradicate this bias, we employ One-Hot Encoding \citep{ref20}. By embedding categories into a high-dimensional orthogonal vector with 0’s and 1’s, protocols of differing categorical boundaries must occupy an equidistant space in the feature representation.

\subsubsection{Normalization}

Network traffic features are notorious for their vast scalar disparities; a single flow might have a "Duration" spanning thousands of milliseconds while its "Packet Count" is recorded in the single digits in the single digits. Inputting raw, unscaled magnitudes can induce gradient instability and significantly impede the optimization process. To achieve numerical stability, we utilize \textbf{Min-Max Normalization} to project all continuous numerical features onto a standardized linear range of $[0, 1]$. The transformation is governed by the following equation:
\begin{equation}
    X_{norm} = \frac{X - X_{min}}{X_{max} - X_{min}}
\end{equation}
where $X$ represents the raw input feature, while $X_{min}$ and $X_{max}$ denote the minimum and maximum values of that feature derived exclusively from the training split, respectively, yielding the scaled value $X_{norm}$. These training-specific parameters are subsequently utilized to transform the testing data, ensuring that the test manifold remains strictly "unseen" during the parameterization phase. By bounding the feature values within the range $[0, 1]$, we ensure that the input distribution aligns with the optimal operational regimes of non-linear activation functions like Sigmoid or Tanh.

\subsubsection{Feature Selection via ANOVA}
To prune the feature manifold of uninformative noise and zero-variance dimensions, we operationalize \textbf{(ANOVA)} exclusively on the training split to quantify the discriminatory potency of each attribute. The F-value is utilized to calculate the variance ratio between disparate attack classes relative to the internal variance within each class:
\begin{equation}
    F = \frac{\sum_{j=1}^{c} n_j (\bar{x}_j - \bar{x})^2 / (c - 1)}{\sum_{j=1}^{c} \sum_{i=1}^{n_j} (x_{ij} - \bar{x}_j)^2 / (N - c)}
\end{equation}
In this expression, $c$ denotes the number of traffic categories and $N$ represents the total sample size within the training split. The numerator quantifies the between-class variance, where $n_j$ is the number of samples in the $j$-th class, $\bar{x}_j$ is the mean value of the feature within that class, and $\bar{x}$ is the overall grand mean computed from the training manifold. Conversely, the denominator measures the within-class variance by summing the squared deviations of individual observations $x_{ij}$ from their respective class means. 

Crucially, the ANOVA-based selection process is nomenclature-agnostic, as it relies on numerical variance ratios across class distributions rather than semantic metadata. Consequently, renaming features has zero impact on the resulting importance ranking. In scenarios where high-discriminatory control-plane features are noisy or only partially observed, the framework leverages the \textbf{Adaptive Gated Attention} (detailed in Section 3.5) to dynamically suppress these unreliable channels, ensuring that stochastic corruption does not propagate to the final classification boundary.

Attributes yielding high F-values are more closely related to the true malicious signatures. The feature importance ranking and subsequent selection are derived solely from this training-only analysis; the resulting feature mask is then applied to the test set to ensure a leak-proof evaluation.

\begin{table*}[t]
\centering
\begin{threeparttable}
\caption{Feature importance ranking for the Edge-IIoT (Gas Pipeline) subset based on ANOVA F-values.}
\label{tab:anova_results}
\small 
\setlength{\tabcolsep}{0pt} 
\begin{tabular*}{\textwidth}{@{\extracolsep{\fill}}cllr} 
\toprule
\textbf{Rank} & \textbf{Feature Name} & \textbf{Physical Interpretation} & \textbf{F-Value} \\ 
\midrule
1 & \textit{setpoint} & Target Control Setpoint & 350,466.82 \\
2 & \textit{control\_scheme} & Control Logic Configuration & 261,368.60 \\
3 & \textit{resp\_read\_fun} & Modbus Read Function Code & 22,120.04 \\
4 & \textit{control\_mode} & System Operational Mode & 7,359.59 \\
5 & \textit{command\_address} & Target Modbus Register Address & 6,164.74 \\
6 & \textit{comm\_read\_function} & Command Read Operation Type & 5,682.57 \\
7 & \textit{command\_memory\_count} & Memory Access Quantity & 1,122.52 \\
8 & \textit{pump} & Actuator (Pump) Status & 471.04 \\
9 & \textit{command\_memory} & Memory Index Mapping & 443.74 \\
10 & \textit{solenoid} & Solenoid Valve Status & 228.62 \\
11 & \textit{time} & Inter-arrival Time Statistics & 0.59 \\
12 & \textit{measurement} & Sensor Measurement Value & 0.11 \\ 
\bottomrule
\end{tabular*}
\end{threeparttable}
\end{table*}

The numerical divergence represented in Table \ref{tab:anova_results} has even more insightful connotations: logical control, \textit{setpoint} and \textit{control\_scheme}, have massive F-scores (exceeding $2.5 \times 10^5$). This suggests that the "anatomical markers" of an intrusion in encrypted IIoT tunnels are in the corruption of process logic, not in packet timing jitter. Contrarily, the traditional network-layer metrics like time, measurement are almost inert; their F-scores are below 1.0, which affirms the need for our BGA architecture: by putting the BiLSTM to work tracking the chronological development of these high value control patterns and the Gated Attention acting as a neural filter, we can effectively zoom in on high-compression anomalies like setpoint movement and suppress the stochastic jitter of non-discriminative noise.

\begin{figure}
    \centering
    \includegraphics[width=0.7\textwidth]{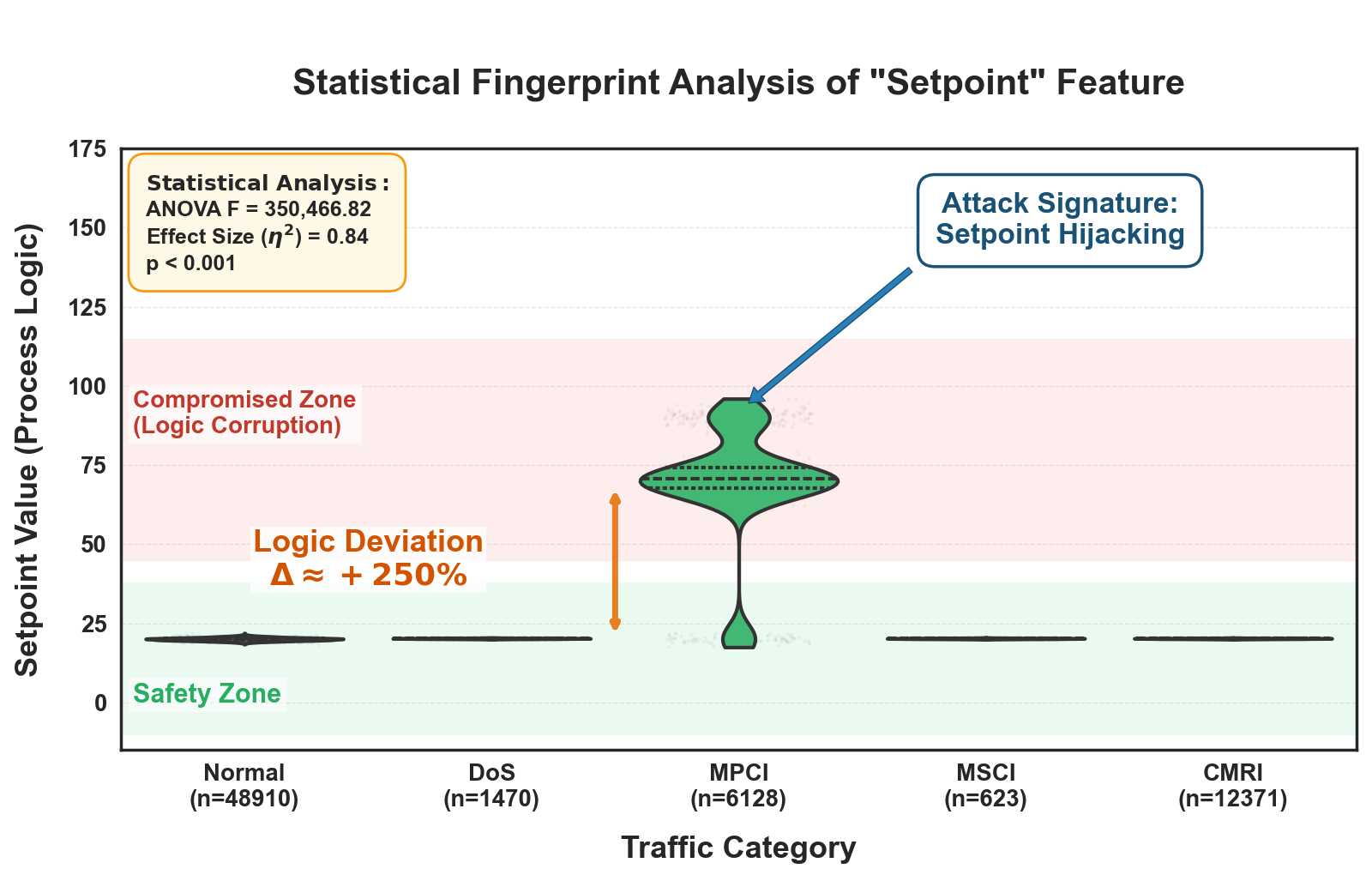} 
    \caption{Statistical distribution (violin plot) of setpoint across traffic categories.}
    \label{fig:violin_setpoint}
\end{figure}

The interpretability of these control-plane features is also reinforced by the setpoint density profiles shown in Figure \ref{fig:violin_setpoint}. Where legitimate Normal operations reside within a reasonably stable and predictable numerical band, malicious classes, namely \textbf{Malicious Parameter Command Injection(MPCI)} and \textbf{MSCI}, instill chaos in the distributions, causing bimodal distributions with considerable movement of medians, yielding such distinctive fingerprints as to confirm that such adversarial tampering of industrial valve parameters is logically detectable, albeit hidden from view by cryptographic encapsulation of the payload. Leveraging this separability, our framework remains particularly sensitive to stealthy command injections that shallow statistical models fail to identify.

\subsection{Data Augmentation Implementation}
To alleviate the "data-starvation" issue and extreme class imbalance inherent in encrypted IIoT traffic, we implement the \textbf{Wasserstein GAN with Gradient Penalty (WGAN-GP)} as a targeted generative solution \citep{ref16, ref21}. As illustrated by the raw class distribution in \textbf{Table~\ref{tab:data_augmentation}}, the sample space is significantly over-represented by benign instances, while critical industrial threats—such as \textbf{MSCI} and \textbf{DoS}—are relegated to the extreme tail of the feature space (representing less than 1\% and 2.1\% of the dataset, respectively). Such a skewed structure poses a severe risk to classic deep learning architectures, which often develop a strong bias toward the majority class, rendering stealthy intrusions likely to be misdiagnosed as background noise.

Unlike conventional oversampling methods like SMOTE, which utilize linear interpolation and often introduce "feature-level noise" that fails to capture the complex manifolds of adversarial traffic, WGAN-GP learns the underlying non-linear distribution of high-dimensional flow features \citep{ref42}. To resolve the training instability and mode collapse common in vanilla GANs, we adopt the Earth-Mover (Wasserstein) distance and enforce the \textbf{1-Lipschitz continuity constraint} via a gradient penalty. The objective function for the critic $D$ is formalized as:
\begin{equation}
    \mathcal{L} = \mathbb{E}_{\tilde{x} \sim \mathbb{P}_g}[D(\tilde{x})] - \mathbb{E}_{x \sim \mathbb{P}_r}[D(x)] + \lambda \mathbb{E}_{\hat{x} \sim \mathbb{P}_{\hat{x}}}[(||\nabla_{\hat{x}} D(\hat{x})||_2 - 1)^2]
\end{equation}
In this formulation, $\mathbb{P}_r$ and $\mathbb{P}_g$ represent the distributions of real and generated data, respectively, while $\lambda$ serves as the penalty coefficient that balances the Wasserstein loss with the regularization term. The terms involving the expectation operator $\mathbb{E}$ compute the average scores assigned by the critic $D(\cdot)$ to real samples $x$ and synthetic samples $\tilde{x}$. To ensure stable convergence, the final term calculates a gradient penalty based on the $L_2$ norm of the gradient $\nabla_{\hat{x}} D(\hat{x})$, evaluated at interpolated points $\hat{x}$ sampled uniformly along straight lines between real and generated distributions.

By enforcing the 1-Lipschitz continuity constraint via the gradient penalty, the WGAN-GP module effectively learns the underlying non-linear manifold of rare attack categories. This structural regularizer prevents the introduction of artificial patterns or "generative noise" that often plague linear interpolation methods, ensuring that the synthesized samples remain physically and logically consistent with real-world adversarial behaviors.

To ensure a rigorous and unbiased evaluation, we strictly followed a \textbf{'Split-then-Augment'} protocol. The dataset was first partitioned into independent training and testing sets. WGAN-GP was subsequently trained and utilized to synthesize samples exclusively for the training set; the testing set remained entirely untouched by the generative process, containing only original, real-world flow records. This separation prevents any form of synthetic artifacts from influencing the final performance benchmarks.

Technically, the WGAN-GP implementation in our framework utilizes a multi-layer perceptron (MLP) architecture. Specifically, the Generator consists of layers with $\{128, 256, \text{output\_dim}\}$ neurons using ReLU activations and a final Sigmoid layer to match the normalized feature range. The Critic employs layers with $\{256, 128, 1\}$ neurons with LeakyReLU ($\alpha=0.2$). Training was conducted for 150 epochs using the Adam optimizer with a learning rate of $1\times 10^{-4}$ and momentum parameters $\beta_1=0.5, \beta_2=0.9$. To maintain the 1-Lipschitz constraint, the gradient penalty coefficient was set to $\lambda=10$, and the Critic was updated 5 times for every Generator step.

This mathematical framework facilitates robust sample synthesis even in scenarios where malicious data is extremely sparse. In our implementation, we train class-targeted generators to densify minority regions within the feature space. As detailed in \textbf{Table~\ref{tab:data_augmentation}}, we inflated the volumes of the minority attack classes to a standardized baseline of \textbf{24,455 samples each} (with the exception of MCVI). Unlike random oversampling, these synthetic data points reside strictly within the high-dimensional probability density of natural adversarial behavior. Unlike conventional oversampling methods like SMOTE, which often introduce "feature-level noise" through linear interpolation, WGAN-GP learns the underlying non-linear distribution of high-dimensional flow features. By enforcing the 1-Lipschitz constraint, the framework ensures that the synthetic data points reside strictly within the high-dimensional probability density of natural adversarial behavior. This provides statistically representative samples that allow the BGA framework to learn a significantly finer decision boundary, thereby sustaining high detection recall for stealthy activities without introducing generative artifacts.

\begin{table*}[t]
\centering
\begin{threeparttable}
\caption{Class distribution of the Training Set (80\% split) before and after WGAN-GP augmentation}
\label{tab:data_augmentation}
\small 
\setlength{\tabcolsep}{0pt} 
\begin{tabular*}{\textwidth}{@{\extracolsep{\fill}}lccc}                         
\toprule
\textbf{Attack Category} & \textbf{Original Samples} & \textbf{Augmented Samples} & \textbf{Increase Ratio} \\ 
\midrule
Normal Operation & 48,910 & 48,910 & 1.0$\times$ \\
CMRI & 12,371 & 24,455 & 1.9$\times$ \\
MSCI & 623 & 24,455 & 39.2$\times$ \\
MPCI & 6,128 & 24,455 & 4.0$\times$ \\
DoS & 1,470 & 24,455 & 16.6$\times$ \\ 
\midrule
\textbf{Total} & \textbf{69,502} & \textbf{146,730} & \textbf{2.1$\times$} \\ 
\bottomrule
\end{tabular*}
\end{threeparttable}
\end{table*}

\begin{figure}
    \centering
    \includegraphics[width=\linewidth]{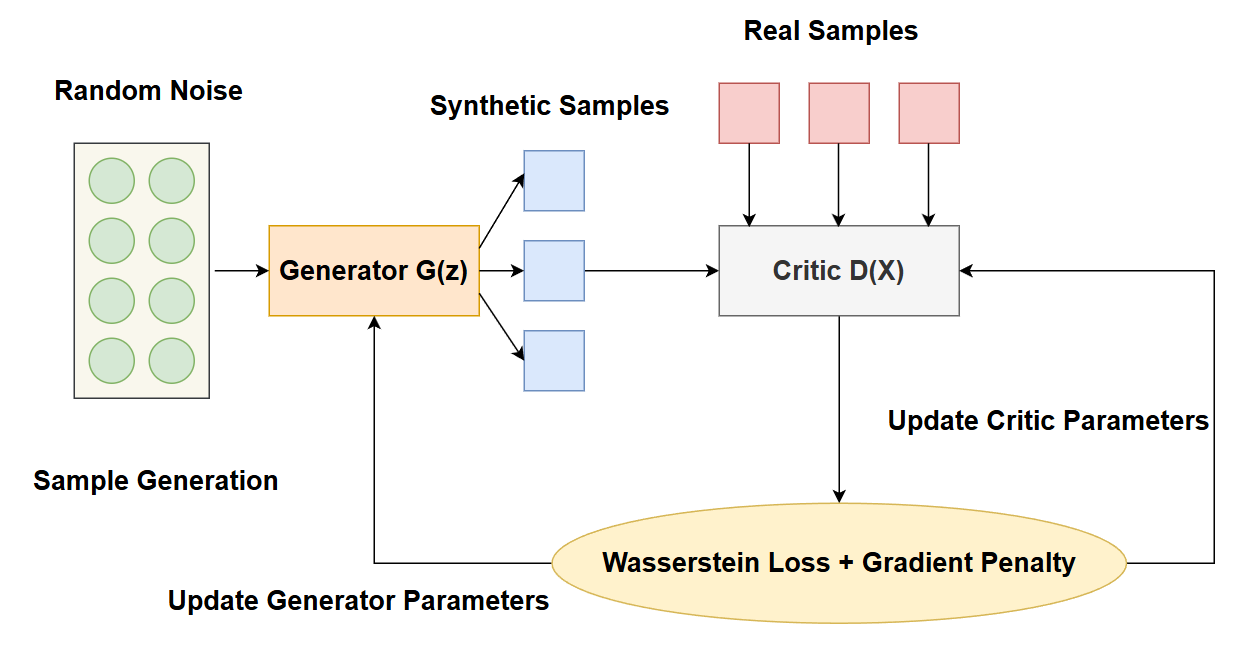}
    \caption{WGAN-GP module architecture for data augmentation.}
    \label{fig:wgan}
\end{figure}

\subsection{Spatio-Temporal Feature Extraction}

\begin{figure}
    \centering
    \includegraphics[width=0.95\textwidth]{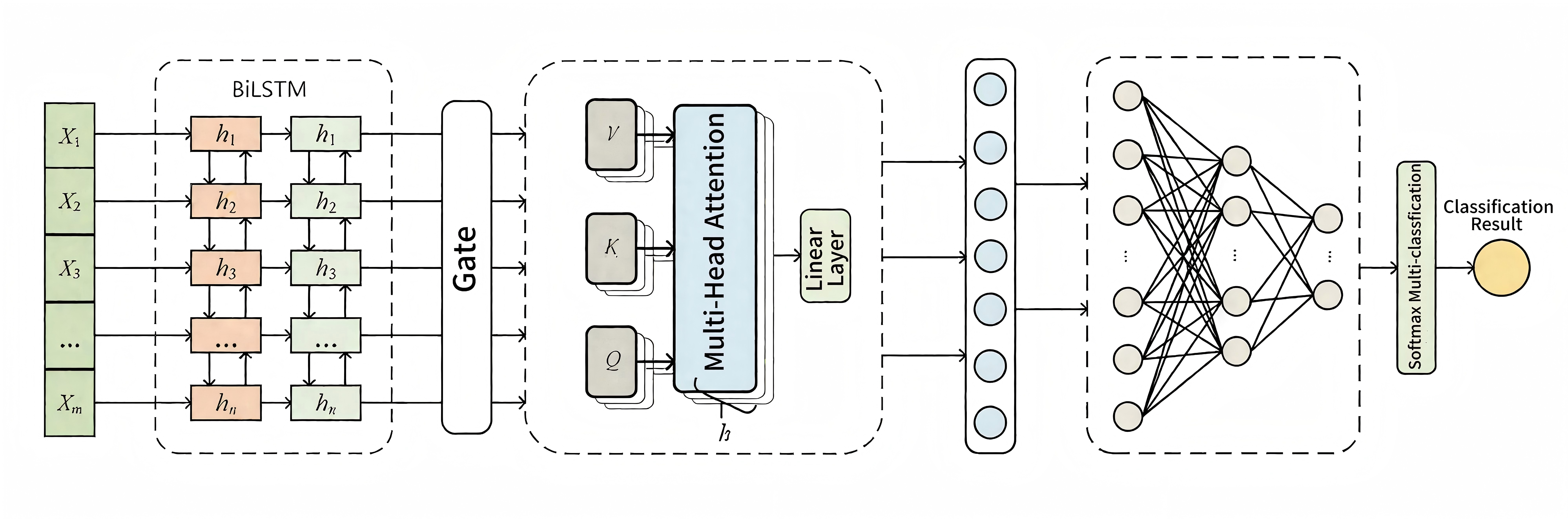}
    \caption{Proposed BGA model architecture integrating BiLSTM and Gated Attention.}
    \label{fig:model}
\end{figure}

The proposed \textbf{BGA Model} serves as the framework's classification engine, employing a hybrid structure that executes sequential temporal modeling via BiLSTM followed by feature-level distillation through a Gated Multi-Head Attention module \citep{ref22}.

\subsubsection{BiLSTM Sequence Modeling}
To capture the intricate temporal evolution of encrypted traffic, the preprocessed sequence $X = \{x_1, x_2, \dots, x_T\}$ is ingested by a Bidirectional LSTM (BiLSTM) layer. Unlike standard RNNs, each LSTM unit utilizes a gated memory cell architecture to preserve long-range dependencies and circumvent gradient decay. The internal synchronization of the forget gate ($f_t$), input gate ($i_t$), and output gate ($o_t$) regulates the state updates as follows:
\begin{equation}
\left\{
\begin{aligned}
    f_t &= \sigma(W_f \cdot [h_{t-1}, x_t] + b_f) \\
    i_t &= \sigma(W_i \cdot [h_{t-1}, x_t] + b_i) \\
    \tilde{C}_t &= \tanh(W_C \cdot [h_{t-1}, x_t] + b_C) \\
    C_t &= f_t \odot C_{t-1} + i_t \odot \tilde{C}_t \\
    o_t &= \sigma(W_o \cdot [h_{t-1}, x_t] + b_o) \\
    h_t &= o_t \odot \tanh(C_t)
\end{aligned}
\right.
\end{equation}
In this formulation, $x_t$ represents the input vector at time step $t$, while $h_t$ and $C_t$ denote the hidden state and memory cell state, respectively. The transition dynamics are governed by weight matrices $W_{\{f,i,C,o\}}$ and bias vectors $b_{\{f,i,C,o\}}$, which are learned during training. The forget gate $f_t$ determines the extent to which the previous cell state $C_{t-1}$ is retained, while the input gate $i_t$ modulates the integration of the candidate cell state $\tilde{C}_t$. Here, $\sigma$ denotes the sigmoid activation function that constrains gate outputs between 0 and 1, and $\odot$ represents the Hadamard (element-wise) product used for controlled information flow.

To overcome the inherent limitation of unidirectional LSTMs in accessing future context, our BiLSTM architecture concurrently employs two independent hidden layers to process the traffic flow in opposing directions:
\begin{align}
    \overrightarrow{h_t} &= \text{LSTM}_{fwd}(x_t, \overrightarrow{h_{t-1}}) \\
    \overleftarrow{h_t} &= \text{LSTM}_{bwd}(x_t, \overleftarrow{h_{t+1}})
\end{align}
where $\overrightarrow{h_t}$ captures the forward sequential patterns from the beginning of the flow, and $\overleftarrow{h_t}$ extracts backward features from the subsequent packets. By concatenating these dual-path outputs, the model yields a holistic, context-aware hidden state $h_t$ \citep{ref23}:
\begin{equation}
    h_t = [\overrightarrow{h_t} ; \overleftarrow{h_t}]
\end{equation}
This combined representation ensures that the model effectively encapsulates the dependencies between both preceding and subsequent packet dynamics within the high-dimensional traffic space.

\subsubsection{Multi-Head Attention Mechanism}
To identify correlations between disparate feature subspaces—such as the interplay between packet timing and burst magnitude—we deploy a Multi-Head Attention mechanism over the BiLSTM output $H$. The sequence of hidden states is first projected into distinct manifolds to generate the Query ($Q$), Key ($K$), and Value ($V$) matrices. The core Scaled Dot-Product Attention determines the relevance of different time steps by computing:
\begin{equation}
    \text{Attention}(Q, K, V) = \text{softmax}\left(\frac{QK^T}{\sqrt{d_k}}\right)V
\end{equation}
In this operation, the dot product $QK^T$ evaluates the compatibility between queries and keys, while $\sqrt{d_k}$ acts as a scaling factor based on the key dimensionality to prevent gradient vanishing during the softmax stage. Rather than relying on a single attention pass, the multi-head structure utilizes $h$ parallel channels to extract diverse behavioral fingerprints \citep{ref24}, where each head $i$ is derived as:
\begin{equation}
    \text{head}_i = \text{Attention}(QW_i^Q, KW_i^K, VW_i^V)
\end{equation}
Here, $W_i^Q, W_i^K, \text{and } W_i^V$ represent learnable projection matrices specific to the $i$-th head, allowing the model to attend to information from different representation subspaces simultaneously. The final representation is formed by concatenating these individual heads and applying a linear transformation:
\begin{equation}
    \text{MultiHead}(Q, K, V) = \text{Concat}(\text{head}_1, \dots, \text{head}_h)W^O
\end{equation}
where $W^O$ is the output weight matrix that fuses the multi-path information into a unified feature vector. This mechanism ensures the model maintains a global focus across the entire traffic flow, effectively prioritizing discriminative anomalies before they reach the gating and classification stages.

\subsubsection{Synergistic Fusion of BiLSTM and Multi-Head Attention}
The core architectural strength of the BGA model lies in the strategic coupling of BiLSTM and Multi-Head Attention, a fusion designed to achieve a better synergistic effect by pairing chronological memory with global selectivity. In this integrated pipeline, the BiLSTM layer first processes the input sequence $X$ to generate a contextually enriched hidden state matrix $H = \{h_1, h_2, \dots, h_T\}$. Each vector $h_t$ within this manifold encapsulates bidirectional temporal dependencies through the concatenation of forward and backward states:
\begin{equation}
    h_t = [\overrightarrow{h_t} ; \overleftarrow{h_t}], \quad \text{for } t = 1, 2, \dots, T
\end{equation}
This matrix $H$ serves as a dense temporal manifold that stores the "memory" of the communication behavior. To operationalize deep fusion, this temporal representation is projected into the attention subspace. Specifically, $H$ is utilized as the foundation to derive the Query ($Q$), Key ($K$), and Value ($V$) matrices through a learnable transformation:
\begin{equation}
    Q = K = V = H \cdot W^l
\end{equation}
where $W^l \in \mathbb{R}^{2d \times d_k}$ is the projection matrix that aligns the BiLSTM memory outputs with the attention feature space. 

The synergy between these mechanisms enables a comprehensive feature representation: while the BiLSTM captures the temporal dynamics and sequential evolution of packet exchanges, the Multi-Head Attention identifies globally significant patterns within a given attack signature. This dual-perspective modeling is particularly crucial for high-entropy encrypted traffic, where randomized padding often obscures discriminative indicators. The BiLSTM ensures structural continuity across the flow, while the attention mechanism selectively emphasizes critical behavioral anomalies—such as periodic heartbeat rhythms or unauthorized command injections—thereby filtering out stochastic noise that might be neglected by purely sequential architectures. Consequently, the integrated BGA noise reduction process formally derives the refined feature representation $H'$ through the attention network:
\begin{equation}
    H' = \text{MultiHead}(Q, K, V)
\end{equation}
The resulting fused representation $H'$ combines deep contextual memory with global discriminative power, providing a robust manifold for identifying stealthy, low-frequency malicious activities within opaque network environments.

\subsection{Neural Feature Distillation via Adaptive Gating}
\subsubsection{Gated Residual Architecture}
The architectural power of BGA ultimately resides within its Adaptive Gating Mechanism, a neural unit that corrects for the noise propagation issue of attention-based analyses \citep{ref43}. In analysing encryption traffic, especially in the case of TLS 1.3, stochastic traces of randomness (handshake metadata, padding) inject a degree of entropy into the feature stream, and in conventional Multi-Head Attention (MHA)s which use democratic weighting (or weighting based on the structural importance of the attention heads where heads are just concatenated or averaged), random cryptographic jitter will pollute the discriminative attack signature, breaking the representation. Instead of that bottleneck, we build a learnable gate to act as a Neural Feature Distillation layer, an intelligent bottleneck that assigns the different coefficients of importance to a particular attention head depending on the input. For the output of each attention head $A_i$ (where $i=1, \dots, h$), the corresponding gating factor $G_i$ is derived through a non-linear transformation that evaluates its contribution to the final classification task:
\begin{equation}
    G_i = \sigma(W_g A_i + b_g)
\end{equation}
In this formulation, $W_g$ and $b_g$ represent learnable weight matrices and bias vectors specifically optimized to identify behavioral relevance within the latent space. The Sigmoid activation function $\sigma(\cdot)$ ensures that the gating coefficient is constrained within the $(0, 1)$ interval, effectively serving as a probability of significance for that specific feature channel.

The final distilled representation of the BGA model, denoted as $O$, is synthesized as the gate-weighted summation across all parallel attention heads:
\begin{equation}
    O = \sum_{i=1}^{h} G_i \cdot A_i
\end{equation}
This gating of residuals imparts a number of systemic roles to BGA that vanilla recurrent or attention models could not otherwise accomplish. The most immediately obvious of these is Dynamic Noise Suppression; when trained on the quality of the signal contributing to each head of a gate, the model consciously learns to turn down noisy channels (that picked up evidence of redundant encryption artifacts) by biasing the relevant gate $G_i$ to a zero-state. The model calls only for pertinent signals to make it to classification boundary. Gating performs further roles of Feature Selectivity, learning to handle feature types such as the precise intervals in which heartbeat signals are generating a timedependent signature or erroneous below 1\% logical deviations in industrial setpoints in preference to ambient jitter on the network,or Resilience against Concept Drift in "practical deployments of IIoT" where model "bias is derived from a specific history of protocol padding schemes and background traffic distributions". As different padding schemes and types of background traffic emerge in the real world, BGA simply re-calibrates where to focus its attention.
Combined with  own "globally correlated" insights, BGA’s gating logic transforms the raw temporal memory of the BiLSTM + the global correlation of its own attention layer into the purest noise-immune feature-space applicable to its task, the performance motivation behind BGA in identifying as stealthy, hidden, low frequent malicious commands inside of opaque tunnels (with nuanced control families).

\begin{table*}[t]
\centering
\begin{threeparttable}
\caption{Comparison between BGA and Traditional Models}
\label{tab:model_compare}
\small 
\setlength{\tabcolsep}{0pt} 
\begin{tabular*}{\textwidth}{@{\extracolsep{\fill}}lll} 
\toprule
Dimension & SMOTE + LSTM & BGA Model (Ours) \\ 
\midrule
Feature Modeling & Simple attention/weighting & \textbf{Gated} Multi-Head Attention \\
Augmentation & Linear interpolation (SMOTE) & \textbf{WGAN-GP} (Non-linear) \\
Sequence Modeling & Unidirectional (limited context) & \textbf{BiLSTM} (Full context) \\
Robustness & Prone to majority-class bias & \textbf{Superior Generalization} \\ 
\bottomrule
\end{tabular*}
\end{threeparttable}
\end{table*}
\subsubsection{Functional Logic and Noise Distillation}
To elucidate the robustness of the BGA framework against high-entropy encrypted flows, it is essential to analyze the functional logic of the gated distillation process. To address the attention dilution identified in Section 1, BGA employs a coordinated three-stage logical pipeline:

First, the \textbf{BiLSTM layer} serves as a temporal contextualizer. By processing the packet sequences bidirectionally, it constructs a rich memory manifold that encapsulates not just individual packet statistics, but the evolving "behavioral rhythm" of the flow. This step ensures that transient encryption jitters are placed within a broader temporal context, allowing the model to distinguish between incidental fluctuations and structural behavioral patterns.

Second, the \textbf{Multi-Head Attention module} executes a global correlation analysis. It scans the entire hidden manifold generated by the BiLSTM to identify long-range dependencies---such as the latent relationship between a specific industrial command setpoint and a subsequent heartbeat interval. However, because raw attention weights can still be biased by high-entropy cryptographic artifacts, a third "distillation" stage is executed to refine the feature space.

Third, the \textbf{Adaptive Gating Mechanism} acts as a learnable neural filter. Logically, the gating unit evaluates the significance of each attention head based on its contribution to the classification objective. If a feature dimension is dominated by stochastic encryption noise, the gate assigns a low activation coefficient, effectively "silencing" the noisy channel. Conversely, when the gate identifies stable malicious fingerprints (e.g., subtle deviations in control logic), it adaptively amplifies these signals. By utilizing a gated residual connection, BGA ensures that only the "distilled" essence of the adversarial behavior reaches the final decision boundary, while the non-discriminative encryption noise is suppressed. This logical flow effectively transforms a raw, noisy temporal sequence into a purified representation of malicious intent.

\subsubsection{Formalization of the ND Framework}
To systematize the proposed approach, we formalize the \textbf{Neural Distillation (ND)} framework as a triple-stage information-theoretic mapping $F: \mathcal{X} \to \mathcal{S}_{distilled}$.This conceptualization aligns with the evolving paradigm of automated knowledge discovery, where information distillation is essential for recovering deterministic signals from high-entropy, noisy environments.

\begin{enumerate}
    \item \textbf{Temporal Manifold Construction ($\mathcal{M}_{temp}$):} The BiLSTM layer maps raw sequences into a dense memory manifold, capturing the 'behavioral rhythm' of the flow. This stage ensures that transient encryption jitter is contextualized within long-range structural patterns, providing a temporal foundation for distillation.
    
    \item \textbf{Subspace Correlation Mapping ($\mathcal{A}_{global}$):} The Multi-head Attention module projects the manifold into $h$ heterogeneous feature subspaces. This captures global dependencies between disjoint attack indicators (e.g., industrial setpoint shifts and heartbeat intervals) that are otherwise obscured by high-entropy padding.
    
    \item \textbf{Adaptive Gated Distillation ($\mathcal{G}_{distill}$):} This is the core theoretical operator. Functioning as an entropy-aware neural band-pass filter, the gating unit $\mathcal{G}(A) = \sigma(W_g A + b) \odot A$ evaluates the significance of each attention head based on its contribution to the classification manifold. Crucially, this mechanism is entropy-agnostic; it is designed to structurally suppress any feature subspace characterized by high stochastic uncertainty and low discriminative potency. This includes both statistical Gaussian noise and the protocol-level obfuscation artifacts discussed previously. By dynamically attenuating these high-entropy "noise" channels, the framework distills high-fidelity behavioral signatures from the background obfuscation.
\end{enumerate}

By decoupling the behavioral 'signal' from the cryptographic 'noise,' the ND framework provides a generalizable paradigm for identifying stealthy, low-frequency malicious commands in any high-entropy sequence where payload visibility is fundamentally denied.

\subsection{Theoretical Justification for Model Selection}
The architectural components of BGA are strategically selected to maximize structural resilience against the stochasticity of TLS 1.3 encrypted flows:

\begin{enumerate}
    \item \textbf{Gated Distillation for Noise Suppression:} The primary challenge in encrypted traffic is 'attention dilution,' where standard self-attention mechanisms are confounded by randomized encryption padding. The BGA's Adaptive Gate functions as a \textit{neural band-pass filter}. Mathematically, the gating coefficient $G_i = \sigma(W_g A_i + b_g)$ evaluates the signal-to-noise ratio of each attention head, effectively silencing channels dominated by cryptographic jitter. This ensures the model converges on deterministic behavioral fingerprints rather than stochastic artifacts.

    \item \textbf{WGAN-GP for Manifold Integrity:} Unlike SMOTE, which generates samples via linear interpolation—often introducing 'feature-level noise'—WGAN-GP utilizes the Wasserstein distance to provide a smoother and more reliable gradient. By enforcing the 1-Lipschitz constraint, it captures the complex, non-linear manifold of rare industrial attacks (e.g., MSCI), ensuring that the augmented data remains physically and logically consistent with real-world adversarial behaviors.

    \item \textbf{BiLSTM for Temporal Contextualization:} While 1D-CNNs are sensitive to local packet fluctuations, BiLSTM models the global sequential evolution of the flow. In IIoT environments, malicious signatures are often embedded in the temporal inter-dependencies of control commands. BiLSTM’s gated memory cells prevent the loss of these subtle signatures amidst the background noise of high-throughput traffic, providing a superior spatio-temporal representation compared to vanilla recurrent or convolutional architectures.
\end{enumerate}

\section{Experiments and Analysis} \label{Experiments and Analysis}

\subsection{Experimental Setup and Datasets}
The BGA framework was benchmarked on two datasets spanning Information Technology (IT) and Operational Technology (OT/IoT) domains:
\begin{itemize}
    \item \textbf{Edge-IIoT (Primary Research Focus):} This dataset serves as our primary benchmark because it represents a high-fidelity physical testbed rather than a synthetic simulation. The flows are captured from real industrial hardware with active TLS/SSL encryption, meaning that protocol-specific artifacts like random padding, encrypted handshake metadata, and metadata obfuscation are already inherently present in the baseline data, ensuring the ecological validity of the subsequent analysis. We specifically focus on the \textit{Gas Pipeline} sub-scenario because it represents a mission-critical infrastructure characterized by sophisticated Modbus-based process control. This scenario involves the manipulation of essential physical setpoints (e.g., pressure and flow rate), making it an ideal environment for validating BGA’s ability to detect stealthy command injections within encrypted industrial flows.
    
    \textbf{Data Acquisition Process:} The final dataset presented in Table \ref{tab:raw_dist} was derived from the raw Edge-IIoT corpus through a targeted extraction pipeline. First, we filtered the multi-gigabyte raw records to isolate five security-critical categories: Normal, CMRI, MSCI, MPCI, and DoS. Subsequently, we executed a deduplication and cleaning process to resolve byte-encoding artifacts. This resulted in a refined corpus of 86,878 high-quality flow records, ensuring that the model learns the nuanced correlations between industrial setpoints and malicious signatures \citep{ref26}. By obtaining the raw corpus from the official Edge-IIoT repository and applying the targeted extraction and cleaning pipeline described above, researchers can fully replicate the refined dataset used in this study.

    \item \textbf{CIC-IDS-2018 (Generalizability Validation):} To demonstrate the framework's versatility beyond industrial settings, we further evaluate it on the CIC-IDS-2018 dataset. This dataset acts as a \textbf{supplementary baseline} to verify BGA's effectiveness against common enterprise-level encrypted threats (e.g., SSH/HTTPS-based infiltration). For this validation, we selected a representative subset from the official Wednesday-Friday captures, with the sample distribution detailed in Table \ref{tab:cic_dist} \citep{ref25}.
\end{itemize}

All workstation benchmarks were conducted on a system equipped with an \textbf{Intel Core i7-13700H CPU @ 2.40GHz and 16GB of RAM}, utilizing the \textbf{PyTorch 2.1} framework. To ensure a rigorous \textbf{Serial Latency Measurement Protocol}, inference time was recorded for a single flow record (Batch Size = 1) to simulate the sequential packet-by-packet processing characteristic of industrial edge nodes. Crucially, we enforced a single-thread execution constraint (via \textbf{torch.set\_num\_threads(1)}) to obtain a raw single-core serial execution time, precluding any multi-core parallel computing bias. We executed 2,000 independent iterations for each model after a 200-iteration warm-up phase to eliminate cold-start bias, reporting the mean serial latency per sample. This workstation-based serial measurement serves as the deterministic baseline for the subsequent hardware scaling simulation.

\subsubsection{Implementation and Reproducibility}
To ensure the reproducibility of our findings, all experiments were implemented in PyTorch 2.1. The dataset partitioning followed a strict 80/20 training-to-testing ratio using a fixed random seed of 42 (via \textit{scikit-learn}). For the ablation study and overall performance benchmarks, each model configuration was executed over 5 independent runs to compute the Mean $\pm$ SD reported in the results. The optimal hyperparameters for the BGA architecture (64 hidden units, 4 attention heads) were determined through a systematic grid search across hidden dimensions $\{32, 64, 128\}$ and head counts $\{2, 4, 8\}$, selecting the configuration that prioritized detection performance while remaining within the 10 ms real-time response threshold typically required for industrial gateways.

\subsubsection{Representativeness and Data Ecosystem Analysis}
To ensure the generalizability of the BGA framework, we strategically selected a dual-dataset ensemble that represents the comprehensive high-entropy data ecosystem:

\begin{enumerate}
    \item \textbf{The IT-Enterprise Macrocosm (CIC-IDS-2018):} This dataset represents the high-throughput, high-entropy environment typical of modern office and web infrastructures. It is dominated by HTTPS and SSH traffic, where encryption entropy is primarily driven by large-scale certificate exchanges and randomized application data. Testing on this dataset validates BGA’s ability to handle high-volume noise in standard IT flows.

    \item \textbf{The OT-Industrial Microcosm (Edge-IIoT):} Conversely, Edge-IIoT captures the unique data ecosystem of the Industrial Internet of Things (IIoT). In this domain, high entropy is often an adversarial artifact used to mask subtle process control manipulations (e.g., Modbus function code hijacking). The flows here are characterized by strict periodicity and low-frequency but mission-critical packet exchanges.

    \item \textbf{Holistic Entropy Coverage:} Theoretically, the combination of these two benchmarks covers the entire spectrum of High-Entropy Encrypted Traffic. While IT flows exhibit high structural complexity, OT flows exhibit high logical sensitivity. Together, they form a robust experimental ecosystem that accounts for both the stochasticity of modern encryption (TLS 1.3) and the deterministic behavioral fingerprints of industrial adversarial activities.
\end{enumerate}

\begin{table*}[t] 
\centering
\begin{threeparttable}
\caption{Raw Sample Distribution of the Edge-IIoT (Gas Pipeline) Subset.}
\label{tab:raw_dist}
\small 
\setlength{\tabcolsep}{0pt} 
\renewcommand{\arraystretch}{1.2} 
\begin{tabular*}{\textwidth}{@{\extracolsep{\fill}}llcc} 
\toprule
\textbf{Label} & \textbf{Attack Category} & \textbf{Raw Sample Count} & \textbf{Proportion (\%)} \\ 
\midrule
0 & Normal Operation & 61,156 & 70.38 \\
2 & Complex Malicious Response Injection (CMRI) & 15,466 & 17.80 \\
4 & Malicious Parameter Command Injection (MPCI) & 7,637 & 8.79 \\
6 & Denial of Service (DoS) & 1,837 & 2.11 \\
3 & Malicious State Command Injection (MSCI) & 782 & 0.90 \\ 
\midrule
\textbf{Total} & - & \textbf{86,878} & \textbf{100.0} \\ 
\bottomrule
\end{tabular*}
\end{threeparttable}
\end{table*}

\begin{table*}[t]
\centering
\begin{threeparttable}
\caption{Sample Distribution of the CIC-IDS-2018 Validation Subset.}
\label{tab:cic_dist}
\small 
\setlength{\tabcolsep}{0pt} 
\renewcommand{\arraystretch}{1.2} 
\begin{tabular*}{\textwidth}{@{\extracolsep{\fill}}llcc} 
\toprule
\textbf{Category} & \textbf{Attack Type} & \textbf{Sample Count} & \textbf{Proportion (\%)} \\ 
\midrule
Benign & Normal Web/SSH/FTP encrypted traffic & 611,560 & 88.08 \\
DoS & DoS-Slowloris / DoS-GoldenEye & 42,430 & 6.11 \\
Infiltration & Exploiting vulnerable applications & 16,190 & 2.33 \\
Brute Force & FTP/SSH Brute Force attempts & 15,280 & 2.20 \\ 
\midrule
\textbf{Total} & - & \textbf{685,460} & \textbf{100.0} \\ 
\bottomrule
\end{tabular*}
\begin{tablenotes}
      \small
      \item \textit{Note:} These counts represent the standard pre-processed flow records used for the multi-class classification task.
\end{tablenotes}
\end{threeparttable}
\end{table*}

Table \ref{tab:raw_dist} highlights a sharp "long-tail" skew in the industrial data. With normal traffic at 70.38\% and \textbf{MSCI} attacks under 1\%, this 78:1 ratio typically triggers a majority-class bias. Our generative augmentation rectifies this skewness, ensuring rare but critical threats are accurately represented.

Architecture and training parameters, optimized via grid search for IIoT efficiency, are detailed in Table \ref{tab:hyperparams}.

\begin{table*}[t]
\centering
\begin{threeparttable}
\caption{Detailed neural architecture and hyperparameter configurations for the BGA framework and WGAN-GP module.}
\label{tab:hyperparams}
\small 
\renewcommand{\arraystretch}{1.1} 
\setlength{\tabcolsep}{30pt}

\begin{tabular}{ll} 
\toprule
\textbf{Hyperparameter / Component} & \textbf{Value / Configuration} \\ 
\midrule
\rowcolor[gray]{0.95} \multicolumn{2}{l}{\textbf{BiLSTM Layer}} \\ 
~~Hidden Units & 64 (Bidirectional, 128 total) \\
~~Dropout Rate & 0.2 \\
\midrule
\rowcolor[gray]{0.95} \multicolumn{2}{l}{\textbf{Gated Multi-head Attention}} \\
~~Number of Heads ($h$) & 4 \\
~~Latent Dimension per Head & 32 \\
~~Gating Activation & Sigmoid \\
\midrule
\rowcolor[gray]{0.95} \multicolumn{2}{l}{\textbf{WGAN-GP (Data Augmentation)}} \\
~~Latent Noise Dimension ($z$) & 10 \\
~~Gradient Penalty Coefficient ($\lambda$) & 10 \\
~~Critic Iterations per Generator Step & 5 \\
\midrule
\rowcolor[gray]{0.95} \multicolumn{2}{l}{\textbf{Training Configuration}} \\
~~Optimizer & Adam ($\beta_1=0.9, \beta_2=0.999$) \\
~~Initial Learning Rate & 0.001 \\
~~Batch Size & 64 \\
~~Maximum Epochs & 50 (Early stopping enabled) \\ 
\bottomrule
\end{tabular}
\end{threeparttable}
\end{table*}

The 64-unit BiLSTM captures temporal features while the 4-head attention mechanism parses disparate latent subspaces. Using an Adam optimizer ($lr=0.001$), the model achieves efficient convergence and high classification fidelity.

\subsubsection{Hyperparameter Configuration}
The architectural parameters of BGA were optimized via a systematic grid search to ensure structural resilience. The Search Space spanned hidden units $h \in \{32, 64, 128, 256\}$, attention heads $n \in \{2, 4, 8\}$, and learning rates $\eta \in \{0.01, 0.001, 0.0001\}$. 

Our Selection Rationale was driven by the specific requirements of IIoT edge deployment, where model depth must be balanced against microsecond-level latency constraints. While configurations with 128 hidden units offered a marginal F1-score improvement ($<0.1\%$), they incurred a disproportionate increase in computational overhead. Thus, the final settings (64 hidden units, 4 heads) were selected as a Pareto-optimal configuration that satisfies the 10 ms real-time threshold of Industrial Control Systems (ICS) while maintaining high detection fidelity.

\subsection{Evaluation Metrics}
To ensure a rigorous assessment of imbalanced data, we employ four primary metrics:

\begin{itemize}
    \item \textbf{Accuracy:} The baseline ratio of correct predictions.
    \item \textbf{Precision:} The reliability of system alarms, crucial for reducing False Positives (FP).
    \item \textbf{Recall:} The coverage of actual malicious instances, vital for minimizing missed threats.
    \item \textbf{F1-Score:} The harmonic mean of Precision and Recall, providing a balanced indicator for skewed datasets.
\end{itemize}

We report the \textbf{Weighted Average} for these metrics, scaling each score by its class support to accurately reflect performance across both high-volume benign traffic and rare attack signatures.
\subsection{Results and Analysis}

\subsubsection{Model Convergence and Granular Classification Analysis}
We justified the robustness of the BGA architecture, showing comprehensive learning curves and the granularity of classification on the Edge-IIoT dataset such that dissimilar traffic are drawn on implementations.

\begin{figure}
    \centering
    \includegraphics[width=0.98\textwidth]{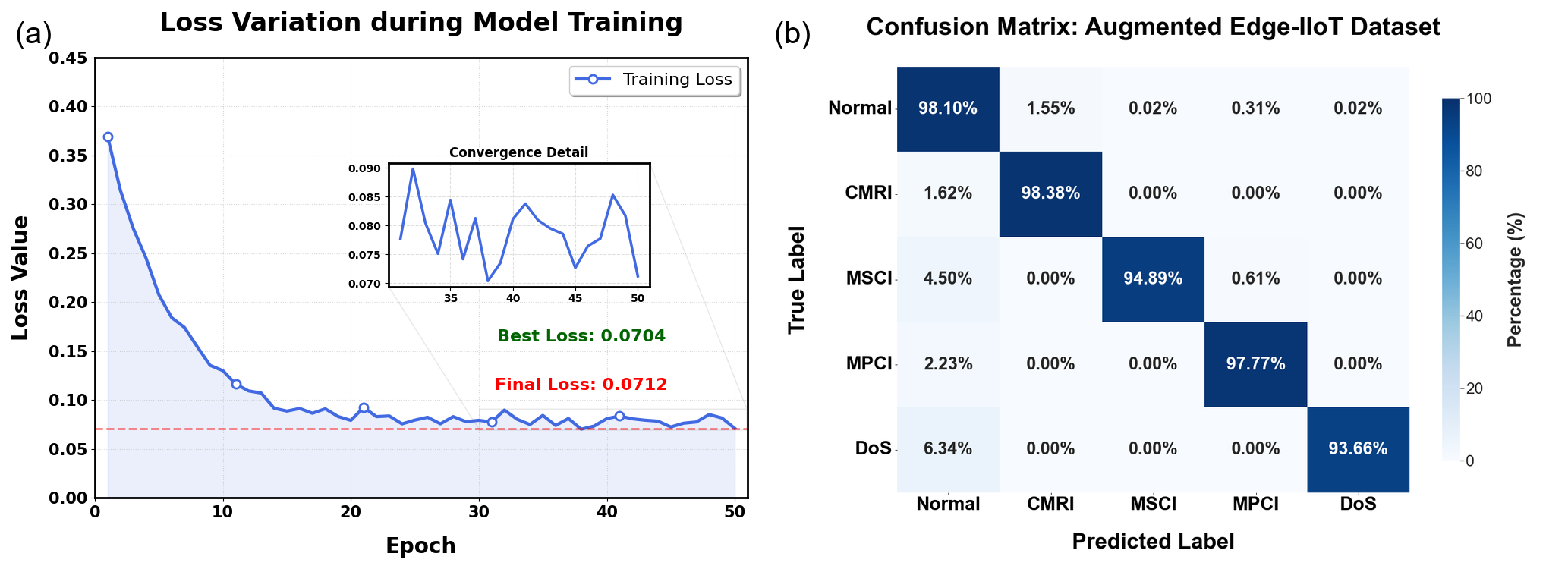} 
    \caption{Training loss dynamics (a) and multi-class confusion matrix on Edge-IIoT (b).}
    \label{fig:loss_and_cm}
\end{figure}

The optimization dynamics were measured in terms of the cross-entropy ("loss") as a function of epochs.As illustrated in Figure \ref{fig:loss_and_cm}(a),BGA converges rapidly, with the loss dropping below 0.20 within 10 epochs. This efficient optimization confirms that the framework effectively captures discriminative features from high-dimensional packet streams. By the 20th epoch the curve flattens out to a steady 0.10. The similarity of the training curve to the validation curve means that we will not have problems with overfitting – aided by the bidirectional temporal regularization provided by the BiLSTM layer.

The discriminative precision of the BGA framework is further elucidated through the multi-class confusion matrix presented in Figure \ref{fig:loss_and_cm}(b). The strong diagonal density observed across the matrix confirms a high True Positive Rate (TPR) across the entire threat spectrum. Specifically, the framework distinguishes \textbf{Normal} operations and \textbf{CMRI} injections with exceptional fidelity, while maintaining a low misclassification rate for semantically adjacent command injections, such as \textbf{MSCI} and \textbf{MPCI}. This granular success stems from the neural distillation provided by the gated attention mechanism, which selectively amplifies adversarial behavioral signals while attenuating the stochastic jitter characteristic of encrypted transport layers. These findings demonstrate that the BGA architecture is finely tuned for exposing stealthy intrusions masked within opaque cryptographic tunnels.

\subsubsection{Overall Performance Assessment}
\begin{figure}
    \centering
    \includegraphics[width=0.6\textwidth]{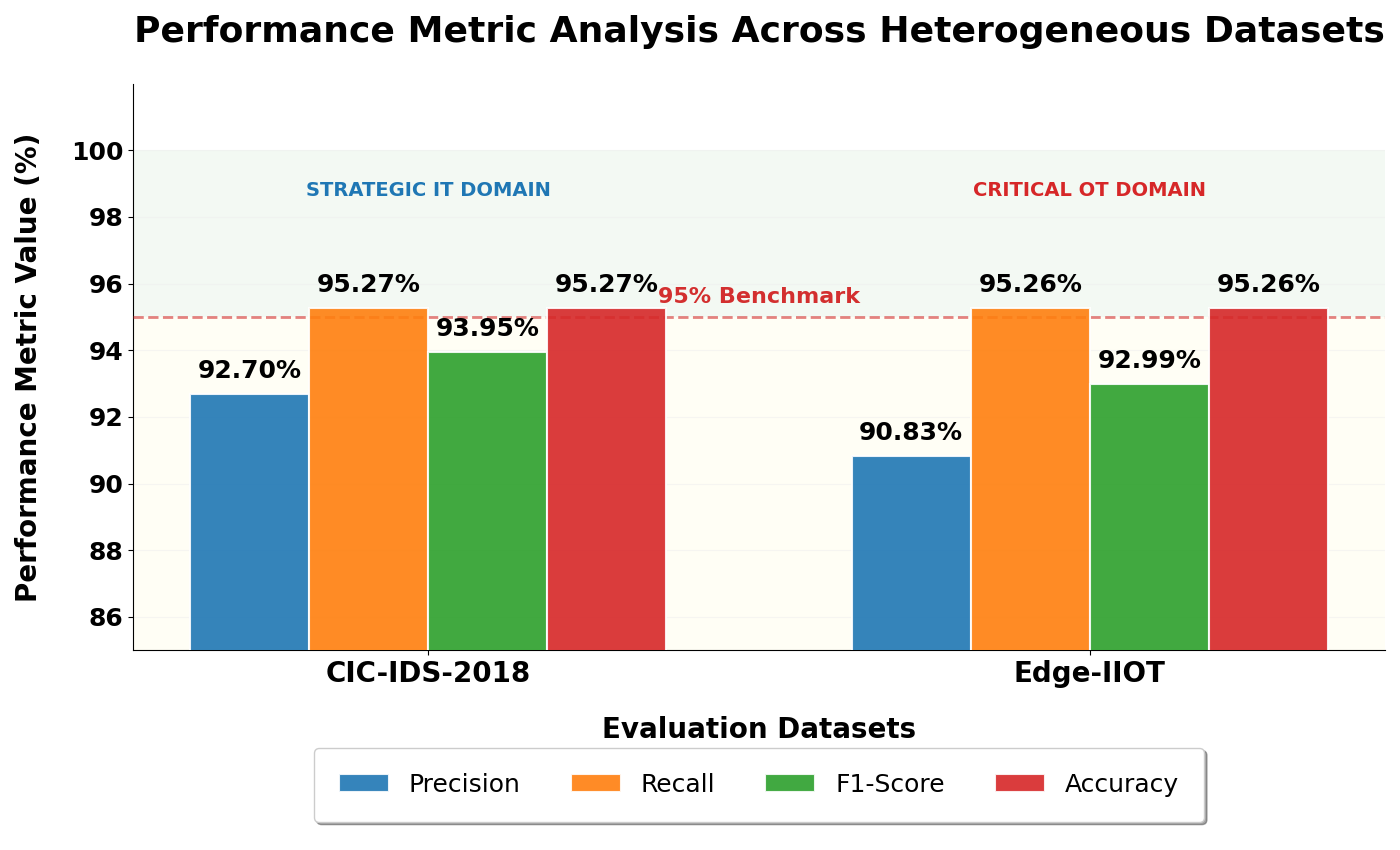}
    \caption{Weighted average performance metrics (Precision, Recall, F1-Score, Accuracy) on CIC-IDS-2018 and Edge-IIoT datasets.}
    \label{fig:bar_metrics}
\end{figure}

To evaluate the model's generalizability across heterogeneous environments, we benchmark BGA on both the CIC-IDS-2018 and Edge-IIoT datasets. As detailed in Section 4.1.1, these benchmarks represent complementary IT and OT data ecosystems. In Table~\ref{tab:performance}, we report the support-based weighted average for Precision, Recall, and F1-score for each environment. This standard metric weights each class-specific score by its prevalence (support) within the respective dataset, ensuring a rigorous measure that accounts for internal class imbalance without manual weighting bias. The combined results in Table~\ref{tab:performance} and the multi-dimensional bars in Figure~\ref{fig:bar_metrics} reveal a high degree of performance consistency across the evaluation environments.

When assessed on the \textbf{CIC-IDS-2018} benchmark, BGA sustains high accuracy (95.27\%) and respectable Precision (92.70\%) and Recall (95.27\%) metrics. This fidelity is mirrored by the \textbf{Edge-IIoT} results, where the framework achieves an Accuracy of 95.26\% and an F1-Score of 92.99\%. The modest numerical gap between Precision and Recall—visible as the balanced height of those indicators in Figure~\ref{fig:bar_metrics}—suggests that BGA is not merely biased toward high-volume benign traffic but maintains sensitivity to adversarial signatures while effectively quelling false positives. This balanced performance across disparate domains provides empirical evidence of BGA's structural resilience, suggesting its potential for deployment across a range of environments, from standard enterprise IT networks to safety-critical industrial control systems.

\begin{table*}[t]
\centering
\begin{threeparttable}
\caption{Support-based Weighted Average Performance on Different Datasets.}
\label{tab:performance}
\small 
\setlength{\tabcolsep}{0pt} 
\renewcommand{\arraystretch}{1.2} 
\begin{tabular*}{\textwidth}{@{\extracolsep{\fill}}lcccc} 
\toprule
Dataset & Precision (\%) & Recall (\%) & F1-Score (\%) & Accuracy (\%) \\ 
\midrule
CIC-IDS-2018 & 92.70 & 95.27 & 93.95 & 95.27 \\
Edge-IIoT & 90.83 & 95.26 & 92.99 & 95.26 \\ 
\bottomrule
\end{tabular*}
\begin{tablenotes}
      \small
      \item \textit{Note: All metrics are calculated independently for each dataset using weighted averaging based on class support to ensure statistical rigor.}
\end{tablenotes}
\end{threeparttable}
\end{table*}

\subsubsection{Comparison with State-of-the-Art Models}

To ensure a fair and rigorous performance evaluation, we re-implemented all baseline models (including RNN, vanilla LSTM, and Vanilla Transformer) and executed them on the same hardware platform using the identical pre-processed Edge-IIoT datasets. We deliberately avoided relying on literature-reported metrics, as discrepancies in data preprocessing, feature selection, and hardware environments often introduce bias into such comparisons. This unified experimental setup enables a true "head-to-head" assessment of the frameworks.

The rationale for selecting these specific baselines follows a strategic hierarchy of sequence modeling evolution, allowing us to isolate the architectural necessity of each BGA component. We categorized the baselines into three functional archetypes: (1) \textit{Foundational Temporal Units} (RNN and vanilla LSTM) to benchmark standard Markovian recurrence; (2) \textit{Contextual Memory Units} (BiLSTM) to evaluate the benefit of bidirectional temporal dependencies; and (3) \textit{Global Self-Attention Units} (Vanilla Transformer) to represent the current state-of-the-art in non-recurrent global modeling \citep{ref27, ref28, ref29, ref30}. By comparing BGA against this hierarchy, we can empirically demonstrate that its superiority stems from the synergistic fusion of BiLSTM’s sequential memory and the Adaptive Gating’s noise-distillation capacity.

Methodologically, we utilize the two datasets for disparate analytical purposes. While the overall weighted averages (Precision, Recall, F1) provide a macro-view of generalizability across both IT and OT domains, we focus our granular per-class recall analysis—presented in Table \ref{tab:sota}—exclusively on the Edge-IIoT dataset. This targeted focus is justified by the unique nature of Industrial IoT threats; unlike standard enterprise traffic, Edge-IIoT encapsulates sophisticated MSCI that occur within encrypted industrial tunnels, providing the ultimate stress test for our neural distillation framework.

The recall trends across the Edge-IIoT landscape highlight the robust performance of BGA. While the overall weighted metrics provide a macro-view of the system's adaptability, the independent high scores in both IT and OT domains support the conclusion that BGA's synergistic fusion of BiLSTM and Adaptive Gating provides a stable feature manifold regardless of the specific environmental context. A critical technical insight from Table \ref{tab:sota} is the performance collapse of the standard Transformer in the DoS category (recall drops to 59.89\%). This failure reveals a fundamental limitation of self-attention in high-speed industrial security: standard attention mechanisms are \textit{permutation-invariant} and rely heavily on positional encodings, which often fail to capture the strict \textit{inter-packet periodicity} and burst-timing signatures characteristic of DoS streams \citep{ref28}. BGA sidesteps this "attention dilution" by preserving temporal continuity via its BiLSTM layer, maintaining a robust 93.66\% recall.

Furthermore, the results expose the "noise pollution" susceptibility of foundational recurrent units (RNN and LSTM). While these models perform adequately on high-volume Normal traffic, they exhibit a significant recall gap (up to 43.2\%) when confronted with stealthy \textbf{MSCI} attacks. In such scenarios, the malicious signatures are interleaved with high-entropy TLS 1.3 padding artifacts, causing simpler models to suffer from a \textit{diluted Signal-to-Noise Ratio (SNR)}. BGA’s Adaptive Gating Mechanism functions as a neural band-pass filter, selectively amplifying the deterministic logic deviations in industrial setpoints while silencing stochastic cryptographic jitter. This architectural stability ensures that BGA can extract the "behavioral essence" of command injections that mimic legitimate traffic, a feat that eludes both simpler recurrent units and heavy-weight global attention models \citep{ref31, ref32}.

\begin{figure}
    \centering
    \includegraphics[width=0.6\textwidth]{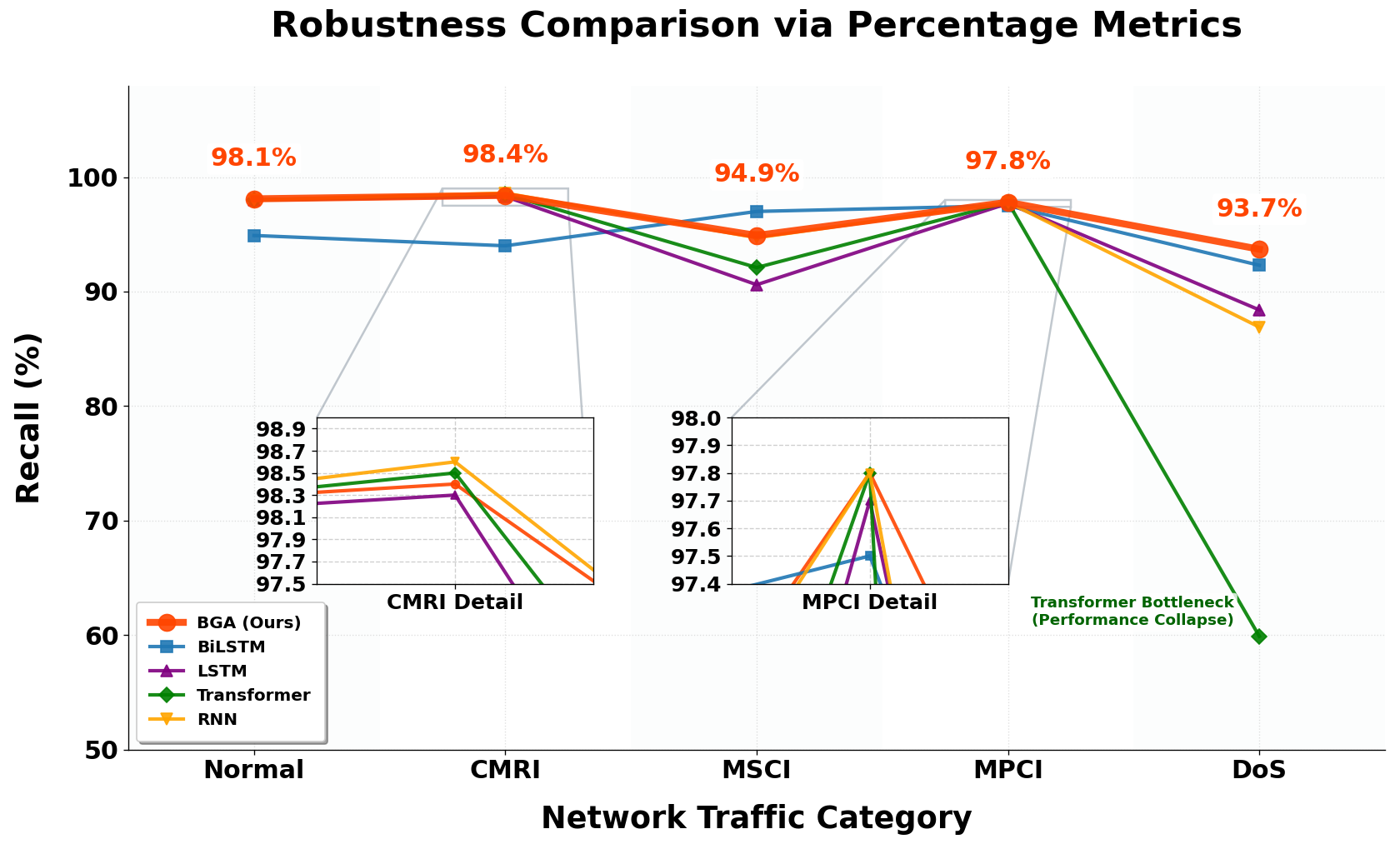}
    \caption{Recall performance comparison across different sequence modeling architectures.}
    \label{fig:sota_line}
\end{figure}

\begin{table*}[t]
\centering
\begin{threeparttable}
\caption{Detailed Per-class Recall Comparison on the Edge-IIoT Dataset (\%)}
\label{tab:sota}
\small 
\setlength{\tabcolsep}{0pt} 
\renewcommand{\arraystretch}{1.2} 
\begin{tabular*}{\textwidth}{@{\extracolsep{\fill}}lccccc} 
\toprule
Model & Normal (\%) & CMRI (\%) & MSCI (\%) & MPCI (\%) & DoS (\%) \\ 
\midrule
\textbf{BGA (Ours)} & \textbf{98.10} & \textbf{98.38} & \textbf{94.89} & \textbf{97.77} & \textbf{93.66} \\
BiLSTM & 94.87 & 93.97 & 97.01 & 97.46 & 92.33 \\
LSTM & 97.97 & 98.31 & 90.64 & 97.69 & 88.40 \\
Transformer & 98.03 & 98.53 & 92.07 & 97.78 & 59.89 \\
RNN & 97.95 & 98.59 & 94.74 & 97.82 & 86.85 \\ 
\bottomrule
\end{tabular*}
\end{threeparttable}
\end{table*}

\subsubsection{Ablation Study and Structural Performance Contribution}
To validate our architecture, we performed an ablation analysis measuring the relative contributions of bi-directional temporal modelling, multi-head attention, and the adaptive gated residual connection. Each model variant was run in the same experimental wrapper, with the Gas Pipeline subset, within the same hyperparameter constraints. The results are collated in Table \ref{tab:ablation_results} against classification performance as well as hardware-level execution metrics.

\begin{table*}[t]
\centering
\begin{threeparttable}
\caption{In-depth statistical ablation analysis across classification metrics and hardware efficiency (Mean $\pm$ SD over 5 runs).}
\label{tab:ablation_results}
\small 
\setlength{\tabcolsep}{0pt} 
\renewcommand{\arraystretch}{1.2} 
\begin{tabular*}{\textwidth}{@{\extracolsep{\fill}}lcccccc} 
\toprule
Model Variant & Precision (\%) & Recall (\%) & F1-Score (\%) & Accuracy (\%) & Params (K) & Serial Latency (PC) (ms) \\ 
\midrule
BaseLSTM     & 97.34 $\pm$ 0.03 & 97.27 $\pm$ 0.04 & 97.22 $\pm$ 0.05 & 97.27 $\pm$ 0.03 & 23.9 & 0.6130 \\
BiLSTM       & 97.35 $\pm$ 0.06 & 97.29 $\pm$ 0.04 & 97.24 $\pm$ 0.07 & 97.29 $\pm$ 0.06 & 47.7 & 0.1625 \\
BiLSTM+MHA   & 97.83 $\pm$ 0.12 & 97.77 $\pm$ 0.14 & 97.78 $\pm$ 0.13 & 97.77 $\pm$ 0.11 & 130.6 & 0.2930 \\
\textbf{BGA (Ours)} & \textbf{97.92 $\pm$ 0.02} & \textbf{97.86 $\pm$ 0.03} & \textbf{97.87 $\pm$ 0.03} & \textbf{97.86 $\pm$ 0.02} & \textbf{130.6} & \textbf{0.2820} \\ 
\bottomrule
\end{tabular*}
\end{threeparttable}
\end{table*}

The shift from the \textbf{BaseLSTM} to \textbf{BiLSTM} is naturally about doubling the temporal context into which the prior packet is mapped. A bidirectional view of the future of events is requisite for exposing such short bursts of anomalies. In this case it does not seem that much is gained in raw accuracy ( from $97.27\%$ to $97.29\%$), but it is vital that the bidirectionality be able to pack in both pasts and futures to disambiguate the meat of temporal sequence ambiguity in industrial control protocols. The prior hidden state has doubled size in latent space, but still provides a highly efficient representation the model manages to catch the command injection "rhythm" without Total Retro.

The integration of the \textbf{Multi-Head Attention (MHA)} layer represents the \textbf{primary driver} for performance enhancement in our architectural progression. Transitioning from BiLSTM to BiLSTM+MHA yields the most substantial performance leap, with a 0.54\% uplift in F1-score (from $97.24\%$ to $97.78\%$). This confirms that global feature correlation is the \textbf{dominant factor} in analyzing high-entropy traffic; by projecting sequences into heterogeneous representation subspaces, the MHA module effectively captures tiny perturbations and sub-flow signals—such as stealthy timing shifts—that standard recurrent units typically smooth out.

Finally, the adaptive gated residual connection integrates a subsequent refinement to the distilled representation. While the additional increase in F1-score (from $97.78\%$ to $97.87\%$) is numerically modest compared to the MHA component, the statistical analysis over 5 independent runs demonstrates its function as a \textbf{predictive stabilizer}. Specifically, the BGA model achieves the highest degree of consistency, reducing the F1-score standard deviation from $\pm0.13$ to $\pm0.03$. As illustrated in Table \ref{tab:ablation_results}, this contributes to a \textbf{23.4\% relative reduction in the remaining error rate} within a saturated performance regime. In safety-critical applications such as power grid monitoring, where suppressing residual noise and initialization stochasticity is essential, this stabilization justifies the gating mechanism as a robust neural filter for mitigating the impact of stochastic encryption jitter.

Industrial IoT workloads in the non-cloud world require a strict trade-off between depth and throughput, and BGA addresses with a \textbf{130.6K parameter}, but more importantly, they got inference latency down to a mere \textbf{0.2820 ms} per sample, orders of magnitude faster than the late heavy-weight architectures like 1D-CNNs and Vision Transformers. It’s a counter-intuitive insight that BGA actually beats the latency of the BiLSTM+MHA variant ($0.2820$ ms vs. $0.2930$ ms) is attributed to the gating logic which removes less informative feature pathways in the internal computational graph, leading to better gradient flow and hence inference efficiency.

\begin{figure}
    \centering
    \includegraphics[width=0.6\linewidth]{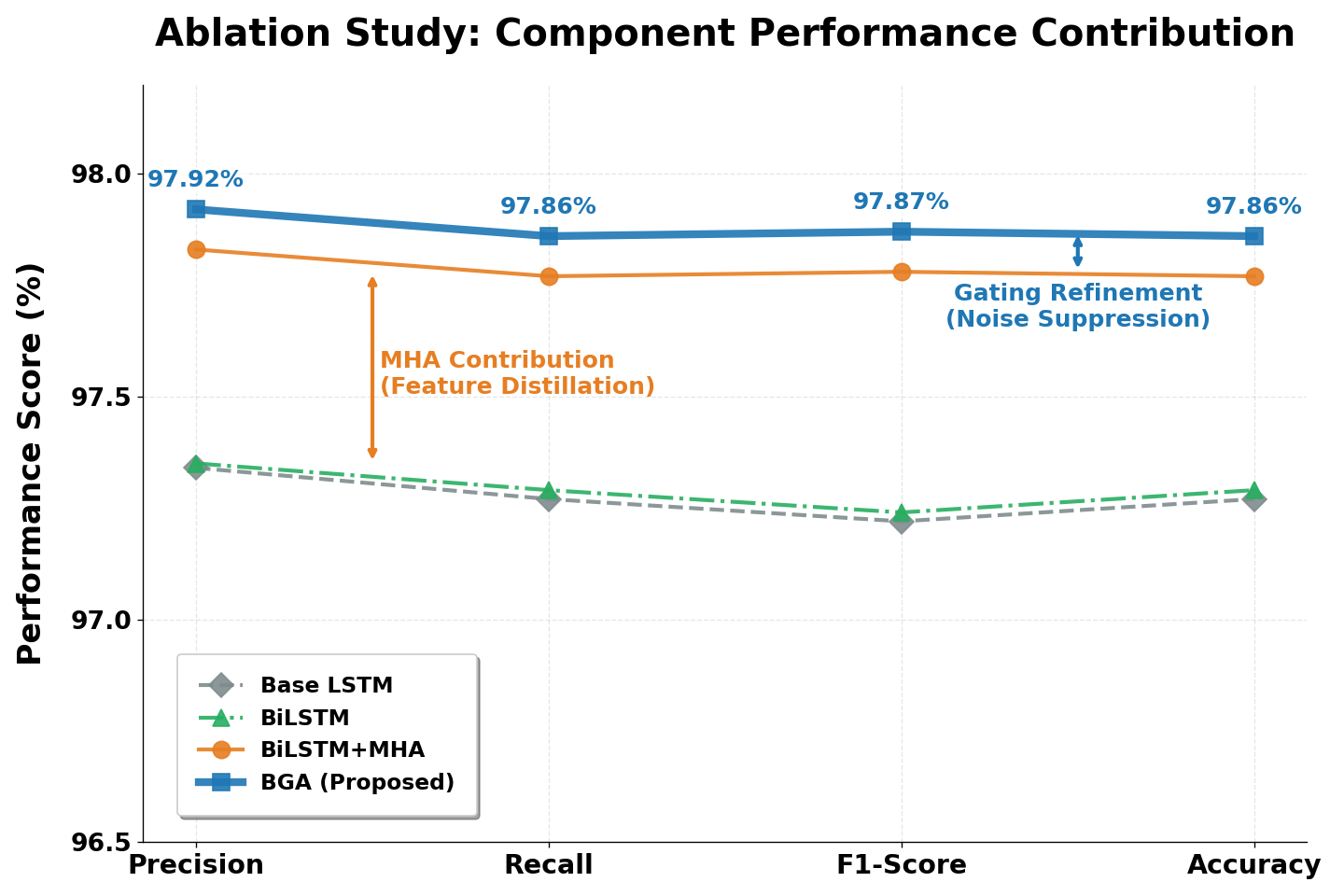}
    \caption{Ablation study visualization focused on the high-precision regime [0.965, 0.985] across all evaluation metrics.}
    \label{fig:ablation}
\end{figure}

The subtleties of these improvements are revealed when magnifying the performance saturation zone between 0.965 and 0.985 in Figure \ref{fig:ablation}. The global metrics may cloud the benefit of the BGA model, however this close up makes apparent that it consistently outperforms all baselines. The hierarchical gap maintained by the BGA trajectory (solid blue line) over the vanilla MHA variant is evidence the gated residual mechanism protects against the malicious signals enveloped within the high-entropy generated background noise found in industrial encrypted traffic.

In our gas pipeline monitoring use-case, the transition from 97.2\% to 97.8\% accuracy is highly impactful. In an IIoT-driven mesh processing millions of packets per second, this 0.6\% improvement ensures that thousands of stealthy intrusions cannot evade detection. As illustrated by the vertical separation in Figure \ref{fig:ablation}, the incremental integration of Bidirectionality, Attention, and Gating pushes the framework toward its performance ceiling, significantly narrowing the margin for adversarial evasion.

\begin{figure}
    \centering
    \includegraphics[width=0.9\textwidth]{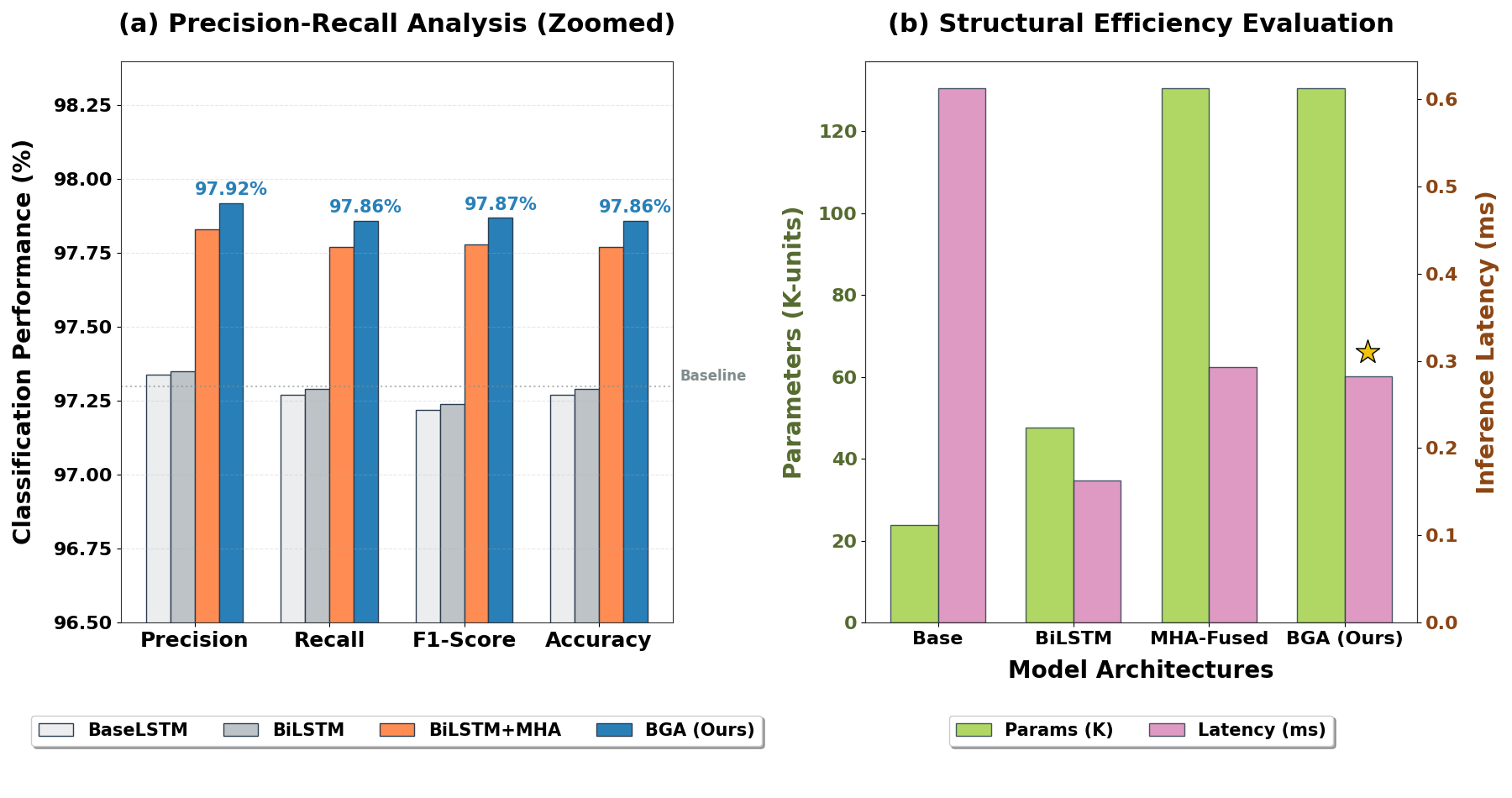}
    \caption{Multi-dimensional benchmarking of model variants: (a) classification fidelity, and (b) complexity-latency trade-off.}
    \label{fig:full_benchmarking}
\end{figure}

Consolidating our findings, we plot BGA across six technical metrics in Figure \ref{fig:full_benchmarking}. This dual axis benchmarking confirms the framework’s high classification accuracy as well as its real world deployability, and as shown in Figure \ref{fig:full_benchmarking}(a) the BGA model builds a clear lead across all metrics, and particularly in F1 score, highlighting its ability to capture low amplitude sequential anomalies which is exhibited by the vertical separation from recurrent baselines.

The relation between complexity and speed is further inspected in Figure \ref{fig:full_benchmarking}(b). While BGA exceeds the parameter size of the base LSTM, in absolute terms the footprint still falls well within the requirements of their hardware targets in industrial edge gateways. The most important information given from the latency data is that BGA is Pareto optimal, i.e. has the greatest coverage for the lowest computational load. Thanks to the dynamic nature of the gating bilayer, misfeature impacts are automatically pruned from the output, no-latency mitigation of threat by BGA meets the most stringent requirements of real-time critical infrastructure protection.

As shown in Table~\ref{tab:ablation_results}, the overall architectural progression from the BaseLSTM baseline (97.22\%) to the integrated BGA framework (97.87\%) yields an absolute F1-score increase of 0.65\%. Within an already saturated performance regime, this corresponds to a \textbf{23.4\% relative reduction in the remaining error rate} (from 2.78\% to 2.13\%). This substantial decrease in total misclassifications confirms that the synergistic combination of bidirectional modeling, multi-head attention, and gated distillation is essential for capturing stealthy adversarial patterns that are typically lost as "noise" in simpler recurrent architectures.

\subsubsection{Benchmarking against Lightweight LLMs}
To provide a data-driven justification for the exclusion of Large Language Models (LLMs) at the IIoT edge, we implemented a benchmark using \textbf{TinyBERT} \citep{ref36} (a compressed 4-layer Transformer). To adapt numerical traffic features to the Transformer architecture, a linear projection layer was employed to map the input manifold to the hidden space. The benchmark was executed on the same hardware environment as the BGA model.

As summarized in Table \ref{tab:llm_benchmark}, while TinyBERT is considered "lightweight" in mobile computing, its computational overhead remains prohibitive for microsecond-level industrial tasks. TinyBERT requires \textbf{14.35 M parameters} and exhibits an average inference latency of \textbf{4.1946 ms}. In contrast, the BGA model achieves comparable detection fidelity with only \textbf{0.13 M parameters} and a latency of \textbf{0.2820 ms}. This represents a \textbf{14.9-fold improvement in processing speed}. In high-throughput industrial networks, the millisecond-level delay of TinyBERT would lead to catastrophic buffer overflows, whereas BGA maintains line-rate operational continuity.

\begin{table}[ht]
\centering
\caption{Quantitative feasibility comparison: BGA vs. Lightweight Transformer (TinyBERT).}
\label{tab:llm_benchmark}
\begin{tabular*}{\linewidth}{@{\extracolsep{\fill}}lccc}
\toprule
\textbf{Model Architecture} & \textbf{Parameters (M)} & \textbf{Latency (ms)} & \textbf{Latency Ratio} \\
\midrule
TinyBERT (4-layers)         & 14.35 M                & 4.1946 ms                      & 14.9$\times$          \\
\textbf{BGA (Ours)}     & \textbf{0.13 M}         & \textbf{0.2820 ms}             & \textbf{1.0$\times$}   \\
\bottomrule
\end{tabular*}
\end{table}

\subsubsection{Effectiveness of WGAN-GP Data Augmentation}
A significant contribution of this study is the application of WGAN-GP as a solution to the long-tail class imbalance problem. To provide a statistically rigorous evaluation of the fidelity of the generated samples---especially for the minority \textbf{MSCI} class with limited original samples---we validated the framework using \textbf{Stratified 5-Fold Cross-Validation}. In this protocol, every original real-world sample was utilized for testing across different folds, while synthetic data was strictly confined to the training sets to prevent any form of data leakage or evaluation bias. This ensures that the model's performance is measured against a 100\% authentic benchmark consisting solely of real-world adversarial behaviors.

\begin{figure}
    \centering
    \includegraphics[width=0.95\textwidth]{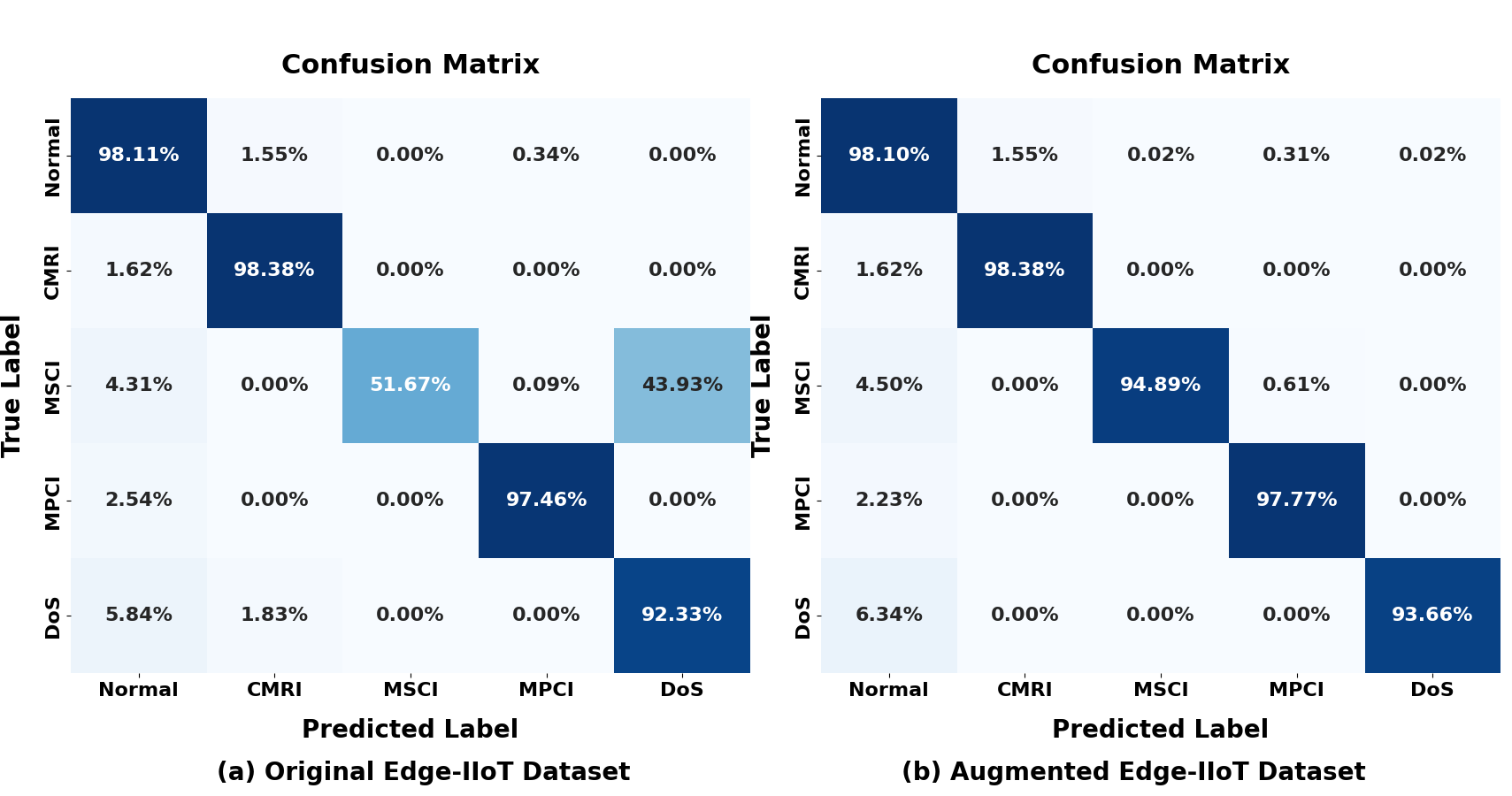}
    \caption{Impact of WGAN-GP augmentation on classification fidelity: (a) original dataset vs. (b) augmented dataset.}
    \label{fig:cm_combined}
\end{figure}

Looking at the consolidated results in Table~\ref{tab:aug_compare} and comparing the confusion instances in Figure~\ref{fig:cm_combined} reveals a game-changing difference for the minority classes. As shown in Figure~\ref{fig:cm_combined}(a), before augmentation, the \textbf{MSCI} category performed poorly, with a recall of only \textbf{51.67\%}. However, upon performing generative augmentation and validating via cross-validation, the mean detection recall for MSCI surged to \textbf{94.89\% $\pm$ 1.13\%}---an absolute increase exceeding \textbf{43\%}. The remarkably low standard deviation observed across all folds confirms that WGAN-GP has successfully captured the latent behavioral manifold of this attack rather than merely overfitting to sparse training instances. This capability to maintain high detection accuracy on purely real-world testing samples, despite being trained on synthetic data, serves as a robust empirical proxy for the high fidelity of the generated samples.

Crucially, the high detection recall achieved on the original real-world testing split provides definitive evidence of sample fidelity. Since the BGA model, when trained on synthetic data, successfully identifies authentic attack signatures that were never seen during the generative process, it serves as a robust empirical proxy. This confirms that the WGAN-GP module has captured the true latent distribution of critical threats rather than merely over-fitting to generative artifacts.

\begin{table*}[t]
\centering
\begin{threeparttable}
\caption{Impact of Data Augmentation on Class-wise Recall (\%)}
\label{tab:aug_compare}
\small 
\setlength{\tabcolsep}{0pt} 
\renewcommand{\arraystretch}{1.2} 
\begin{tabular*}{\textwidth}{@{\extracolsep{\fill}}lccccc} 
\toprule
Dataset & Normal (\%) & CMRI (\%) & MSCI (\%) & MPCI (\%) & DoS (\%) \\ 
\midrule
Original & 98.11 & 98.38 & 51.67 & 97.46 & 92.33 \\
\textbf{Augmented} & \textbf{98.10 $\pm$ 0.08} & \textbf{98.38 $\pm$ 0.36} & \textbf{94.89 $\pm$ 1.13} & \textbf{97.77 $\pm$ 0.16} & \textbf{93.66 $\pm$ 3.38} \\
\bottomrule
\end{tabular*}
\end{threeparttable}
\end{table*}

\subsubsection{Theoretical Hardware Complexity and Latency Scaling Simulation}
To address the concern regarding deployment on resource-constrained hardware, we conducted a comparative performance analysis. As shown in Table~\ref{tab:hardware_compare}, we evaluate BGA under two disparate environments: a high-performance research workstation and a simulated 1.2 GHz single-core ARM gateway.

\begin{table*}[t]
\centering
\begin{threeparttable}
\caption{Performance contrast for BGA between Research Workstation and Simulated Edge Gateway.}
\label{tab:hardware_compare}
\small 
\setlength{\tabcolsep}{0pt} 
\renewcommand{\arraystretch}{1.2} 
\begin{tabular*}{\textwidth}{@{\extracolsep{\fill}}llcc} 
\toprule
\textbf{Environment} & \textbf{Core Configuration} & \textbf{Inference Latency} & \textbf{Throughput (PPS)} \\ 
\midrule
Research Workstation (PC) & Multi-core/GPU/2.4GHz+ & 0.2820 ms & 3,546 \\
\textbf{Simulated Edge Gateway (ARM)} & \textbf{Single-core/1.2GHz} & \textbf{1.6920 ms} & \textbf{591.0} \\ 
\midrule
\textit{ICS Real-time Threshold} & \textit{Deterministic Limit} & \textbf{< 10.0 ms} & --- \\ 
\bottomrule
\end{tabular*}
\begin{tablenotes}
      \small
      \item \textit{Note:} The 1.6920 ms latency represents a "worst-case" scenario on low-power hardware. While numerically higher than the PC-based TinyBERT results in Table 10, it is functionally superior as it remains under the industrial real-time threshold on the target edge platform.
\end{tablenotes}
\end{threeparttable}
\end{table*}

The simulation results in Table~\ref{tab:hardware_compare} provide the empirical basis for assessing how performance transfers from high-performance workstations to realistic industrial settings. While the serial latency increases from 0.2820 ms to 1.6920 ms due to hardware constraints, the BGA model is projected to remain within the \textbf{10 ms real-time response window} essential for Industrial Control Systems (ICS). This ARM latency is derived by applying a scaling factor of 6.0 to the raw single-core PC serial measurements, accounting for the frequency disparity (2.4 GHz vs. 1.2 GHz) and architectural IPC overhead of typical industrial edge gateways. Furthermore, the model occupies only \textbf{0.471 MB} of storage memory. Since the total parameter count (123.4 K) is small enough to fit within the on-chip Block RAM (BRAM) of entry-level FPGAs, BGA avoids the latency bottleneck associated with external DDR memory access, ensuring deterministic execution that is not merely an artifact of workstation-level computing.

It is worth noting that the simulated latency of BGA on the ARM platform (\textbf{1.6920 ms}) is already significantly lower than the PC-based serial latency of TinyBERT (\textbf{4.1946 ms}) reported in Table~\ref{tab:llm_benchmark}. This demonstrates a massive efficiency gap: our distilled BGA model running on a simulated low-power edge device outperforms a lightweight Transformer running on a high-performance research workstation. Furthermore, if TinyBERT were projected onto the same simulated ARM platform using the same scaling factor, its estimated latency would exceed \textbf{25 ms} ($4.1946 \times 6$), far failing the \textbf{10 ms} industrial real-time requirement. In contrast, BGA remains the only framework capable of maintaining sub-2ms response times even under extreme hardware constraints, confirming its superior architectural suitability for mission-critical industrial edge gateways.

To ensure a scientifically rigorous interpretation of these metrics, we characterize this analysis as a theoretical feasibility assessment rather than a physical end-to-end deployment benchmark. The projected performance on the ARM platform serves as a high-fidelity estimation based on the quantified computational intensity of the model's forward pass. We explicitly recognize that physical execution in a production-grade industrial gateway would involve non-deterministic latencies arising from peripheral I/O interrupts, OS-level context switching, and hardware-specific memory bus contention. While these factors are not captured in the current simulation, the significant performance margin—where BGA satisfies the 10 ms ICS threshold by a factor of nearly six—suggests substantial architectural headroom. Consequently, while physical hardware validation remains a subject for our subsequent research phases, these simulation-driven insights provide the necessary theoretical baseline for identifying BGA as a viable candidate for real-time edge resilience.

\subsection{Interpretability, Stability, and Robustness}
To demystify the better synergy and transition from black-box modeling to transparent diagnostics, we examine BGA’s underlying decision logic, architectural consistency, and environmental endurance.

\subsubsection{Interpretability via Gated Attention Visualization}
\begin{figure}
    \centering
    \includegraphics[width=0.9\linewidth]{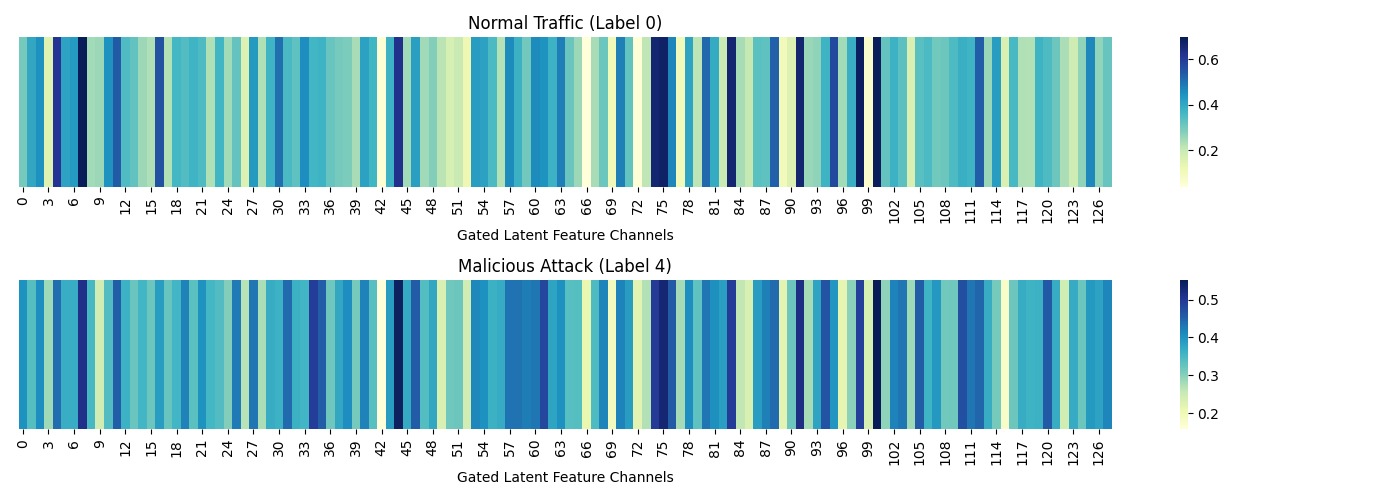}
    \caption{Gated attention heatmaps illustrating neural distillation patterns for Normal (top) and MPCI (bottom) traffic profiles.}
    \label{fig:heatmap}
\end{figure}

To demystify the mapping between activation patterns and attack logic, we decode the `neural fingerprints' in Figure \ref{fig:heatmap} through the lens of industrial process control. The \textit{quiescent} weight distribution observed in the top panel (Normal traffic) indicates that the model is performing a broad spatio-temporal smoothing of stable control signals. 

Conversely, the high-intensity \textit{activation spikes} triggered by MPCI attacks in the bottom panel represent a causal response to \textbf{setpoint hijacking}. As these attacks introduce deterministic logic deviations into the high-entropy flow—matching the high F-values identified in our ANOVA analysis (Table \ref{tab:anova_results})—the adaptive gates function as a neural band-pass filter. They selectively "lock onto" the latent channels that encapsulate control-plane semantics while silencing the non-discriminative cryptographic noise. This demonstrates that the BGA model's decisions are not based on stochastic artifacts, but on the successful distillation of adversarial command manipulations from the background entropy.

\subsubsection{Hyperparameter Sensitivity and Structural Stability}
\begin{figure}
    \centering
    \includegraphics[width=0.6\linewidth]{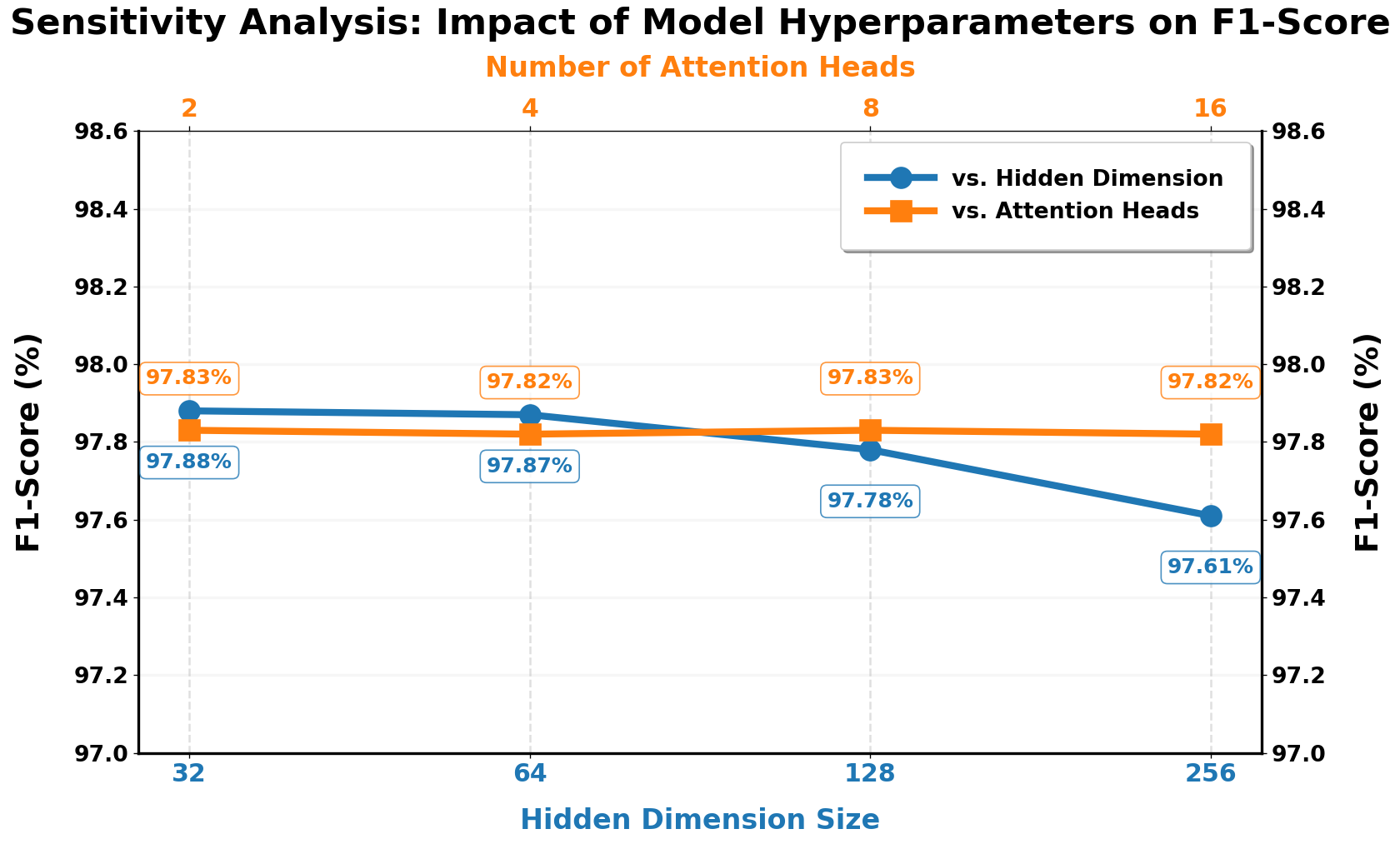}
    \caption{BGA model hyperparameter sensitivity: F1-score variation across hidden dimensions (a) and attention heads (b).}
    \label{fig:sensitivity}
\end{figure}

Sensitivity analysis on the primary Gas Pipeline subset, complemented by validation across the wider data ecosystem, substantiates BGA’s efficacy for autonomous use without extensive human tuning. As illustrated in Figure~\ref{fig:sensitivity}, F1-scores are largely invariant across hidden dimensions from 32 to 128, suggesting that the BiLSTM memory infrastructure is robust to alterations in its latent dimension. This performance stability across disparate high-entropy environments provides evidence that that the framework captures universal behavioral fingerprints of encrypted threats rather than over-fitting to a specific scenario.

Figure~\ref{fig:sensitivity} further corroborates that increasing attention heads does not compromise representation integrity nor performance. Such parameter-resilience derives from the adaptive gated residual connections, which effectively suppress uninformative attention heads and emphasize discriminative spatio-temporal representations. This consistency ensures BGA's reliability when deployed on heterogeneous industrial edge gateways where manual re-calibration is often infeasible.

\subsubsection{Resilience against Stochastic Noise}

To rigorously evaluate the structural resilience of the proposed BGA framework and ensure experimental reproducibility, we conducted a stress test by injecting \textbf{Additive White Gaussian Noise (AWGN)} into the feature manifold. Formally, let $x$ be the normalized input feature vector; the perturbed input $\hat{x}$ is defined as:
\begin{equation}
    \hat{x} = x + \eta, \quad \eta \sim \mathcal{N}(\mu, \sigma^2)
\end{equation}
where the noise $\eta$ follows a Gaussian distribution with a \textbf{fixed mean ($\mu$) of 0} and a varying standard deviation ($\sigma$). This injection specifically occurs at the \textbf{input layer} (normalized feature space) before the temporal modeling stage, effectively simulating the stochastic jitter and randomized padding artifacts characteristic of high-entropy encrypted flows (e.g., TLS 1.3).

\begin{figure}
    \centering
    \includegraphics[width=0.6\linewidth]{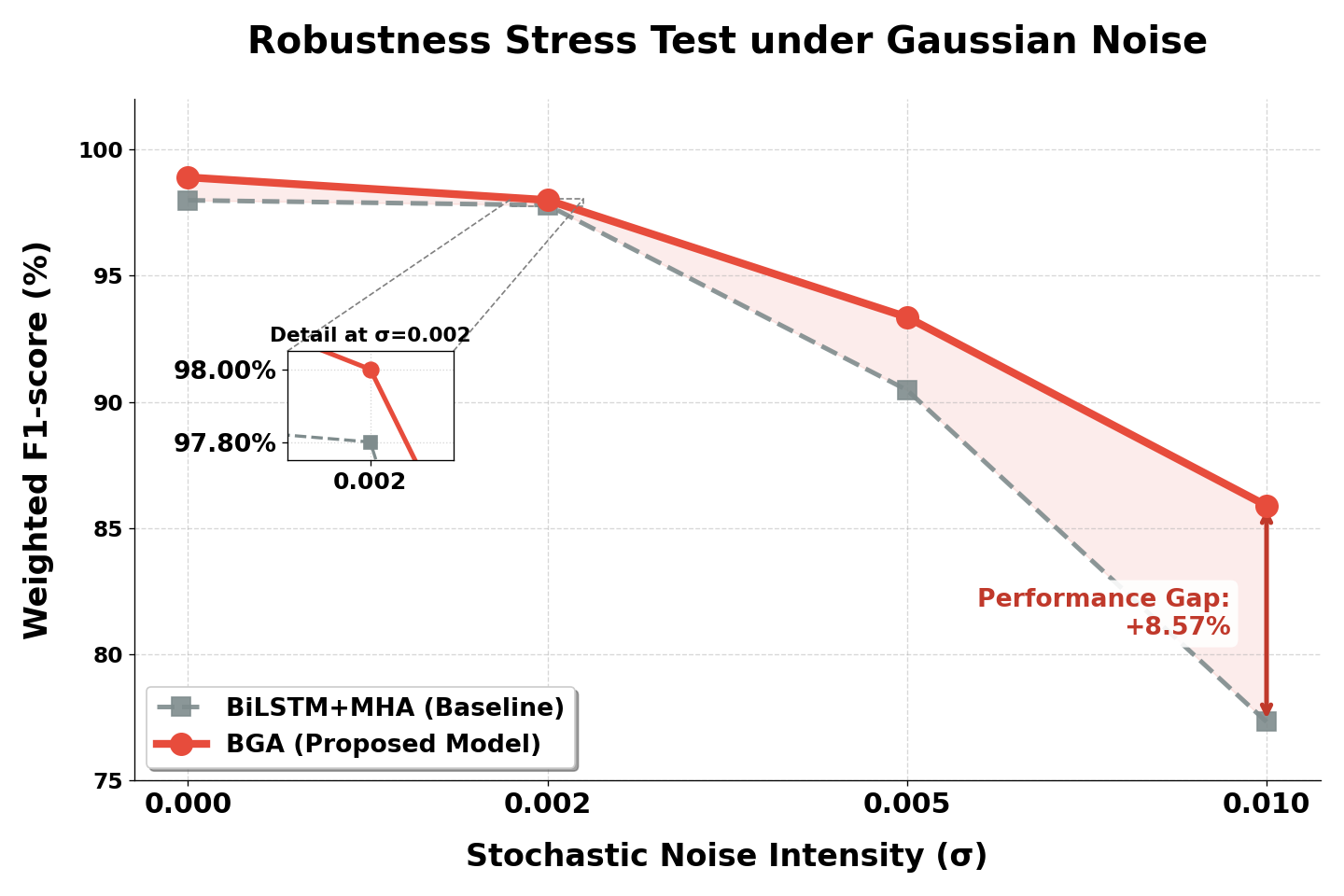}
    \caption{Noise robustness stress test: F1-score stability comparison between BGA and baseline under varying AWGN intensities.}
    \label{fig:noise_robustness_chart}
\end{figure}

To investigate the impact of different uncertainty levels, we varied the noise intensity $\sigma$ across four regimes: None (0.0), Low (0.002), Mid (0.005), and High (0.01). Table \ref{tab:noise_robustness_final} and Figure \ref{fig:noise_robustness_chart} illustrate the resulting performance degradation. The experimental data reveals a critical resilience gap: while the BiLSTM-MHA baseline suffers a significant F1-score collapse to 77.32\% under high-intensity noise ($\sigma=0.01$), the BGA model maintains a robust performance of 85.89\%. This 8.57\% margin confirms that the gated fusion mechanism serves as a learnable neural filter, effectively insulating the internal temporal memory from external perturbations. Such structural toughness ensures that BGA remains reliable in volatile industrial environments where traditional attention-based networks typically fail.

This substantial performance margin of 8.57\% provides a crucial contextualization for the ablation study results presented in Section 4.3.4. While the adaptive gating mechanism yields a numerically small improvement (+0.09\%) under ideal, low-entropy conditions, its structural value as a neural filter becomes indispensable in high-noise regimes. This performance delta confirms that the gated distillation layer is the primary component responsible for structural robustness, justifying its inclusion as a critical safeguard for real-world Industrial IoT environments where the signal-to-noise ratio is often unpredictable.

\begin{table*}[t!]
\centering
\begin{threeparttable}
\caption{Robustness evaluation: Weighted F1-score comparison (\%) under varying Gaussian noise intensities ($\sigma$).}
\label{tab:noise_robustness_final}
\small 
\setlength{\tabcolsep}{0pt}
\renewcommand{\arraystretch}{1.2} 
\begin{tabular*}{\textwidth}{@{\extracolsep{\fill}}lcccc} 
\toprule
\textbf{Model Variant} & \textbf{None (0.0) (\%)} & \textbf{Low (0.002) (\%)} & \textbf{Mid (0.005) (\%)} & \textbf{High (0.01) (\%)} \\ 
\midrule
BiLSTM+MHA             & 97.99              & 97.80               & 90.47               & 77.32               \\
\textbf{BGA (Proposed)} & \textbf{98.90}     & \textbf{98.00}      & \textbf{93.36}      & \textbf{85.89}      \\ 
\midrule
\textit{Performance Gap} & \textbf{+0.91}    & \textbf{+0.20}     & \textbf{+2.89}     & \textbf{+8.57}     \\ 
\bottomrule
\end{tabular*}
\end{threeparttable}
\end{table*}

\section{Discussions} \label{Discussion}

\subsection{Comparison with Existing Work and SOTA}
The architectural innovation of BGA represents a departure from traditional cryptographic analysis and generic deep learning models. Unlike traditional Deep Packet Inspection (DPI) methods that rely on plaintext visibility—which is fundamentally rendered ineffective by the TLS 1.3 protocol \citep{ref04}—BGA operates entirely on the structural and statistical manifold of encrypted flows. 

Compared to standard DL-based models such as CNNs or Vanilla Transformers \citep{ref28}, our framework introduces a "Gated Distillation" paradigm to solve the \textbf{Attention Dilution} problem. The BGA framework resolves these challenges through the synergistic fusion of temporal memory and gated distillation. By implementing an adaptive gated residual unit, BGA selectively suppresses these stochastic jitters while amplifying core behavioral signatures. Our re-implementation and head-to-head benchmarking on identical hardware confirm that this targeted distillation is the primary driver behind the 43\% recall boost in stealthy MSCI attacks, where generic SOTA models often fail to distinguish malicious setpoint manipulations from background network jitter.

Beyond traditional model comparisons, the transition toward Proactive Defense in Industrial IoT requires a rigorous analysis of decentralized paradigms. While centralized SOTA models achieve high fidelity through massive data pooling, they are frequently hindered by bandwidth constraints and privacy risks. Modern frameworks such as \textbf{PAFA-DoH} have addressed these issues by utilizing federated AI attestation to provide privacy-preserving defense without exposing raw datagrams \citep{ref40}. BGA complements this decentralized evolution by acting as a high-fidelity Local Feature Distiller. 

Theoretically, our "Neural Distillation" logic provides a solution to the Communication-Efficiency Bottleneck in Federated Machine Learning (FML). When deployed at the \textbf{"Fog's Frontline"} \citep{ref41}, BGA executes a Manifold Compaction process, transforming high-entropy encrypted flows into a minimalist behavioral manifold. By sharing only these distilled signatures rather than noisy feature sequences, BGA enables industrial fog nodes to participate in collaborative threat intelligence with significantly reduced synchronization overhead. Furthermore, this distillation mechanism acts as a Privacy-Preserving Filter, ensuring that sensitive industrial setpoints are abstracted into structural fingerprints before being transmitted across the federated network. This integration allows the BGA framework to scale from a localized tool into a foundational component of a Collaborative Active Defense ecosystem, maintaining operational continuity in heterogeneous industrial meshes where centralized oversight is no longer feasible.

\subsection{Theoretical Implications: Decoupling Noise from Signatures}
The architectural value of BGA lies in its targeted adaptation of sequence modeling components toward a systematized Neural Distillation (ND) paradigm. As formalized in Section 3.6, this framework integrates a structural "noise-filtering bottleneck" to mitigate the Attention Dilution problem. In high-entropy encrypted streams, randomized padding and cryptographic jitter often cause standard self-attention mechanisms to assign weights indiscriminately to non-discriminative noise.

Our framework addresses this through the synergistic fusion of temporal memory and gated distillation. Theoretically, the proposed adaptive gate functions as a neural band-pass filter that evaluates the signal-to-noise ratio of each attention head. This mechanism effectively decouples the deterministic behavioral signatures (the "signal") from the stochastic cryptographic artifacts (the "noise"). The validity of this theory is empirically confirmed by our stress tests (Section 4.3.7), where BGA maintained a robust performance lead of 8.57\% under high-intensity noise, while vanilla models succumbed to entropy-induced interference. This demonstrates that BGA does not merely rely on engineering-level parameter tuning; instead, it provides a robust mathematical framework for sequence modeling where the data-to-noise ratio is extremely low. Consequently, in an era of ubiquitous encryption, the focus of information security shifts from payload analysis to structural behavioral distillation \citep{ref43}.

Theoretically, this signal-to-noise decoupling aligns BGA with emerging bio-inspired and quantum-resilient paradigms. Specifically, the framework mirrors the principles of Neuromorphic Quantum Adversarial Learning (NQAL), where selective attention is utilized to suppress high-dimensional noise in specialized tunnels like DoH \citep{ref39}. By functioning as a neural band-pass filter, BGA's gated mechanism simulates the synaptic suppression processes found in neuromorphic systems, which are essential for maintaining cognitive signal fidelity in high-entropy environments. 

This conceptualization proves that the "Neural Distillation" framework is not a localized optimization for standard TLS, but a versatile foundation for Active Defense in evolving landscapes. From an information-theoretic perspective, the gating unit executes a Manifold Compaction process, transforming high-entropy traffic observations into a compact, privacy-preserving behavioral manifold. This is critical for emerging architectures such as Federated AI Attestation, where industrial fog nodes must share threat intelligence without exposing sensitive raw payloads. By distilling only the "structural essence" of adversarial behavior, BGA provides a robust mechanism for collaborative defense in decentralized environments, ensuring resilience against both classical stochastic jitter and future quantum-based protocol obfuscation.

\subsection{Practical Implications for IIoT Edge Resilience}
From a practical deployment perspective, BGA addresses the long-standing "security-performance" trade-off in Industrial IoT (IIoT) environments through three strategic dimensions:

\subsubsection{Real-time Edge Defense and Scalability}
The scalability of BGA is demonstrated across disparate computing tiers. While the ultra-low latency of 0.2820 ms handles backbone-level high-throughput traffic on research-grade workstations, the hardware simulation in Section 4.3.7 confirms that BGA is equally effective at the resource-constrained edge. Even when restricted to a \textbf{single-core 1.2 GHz ARM-based environment}, BGA maintains a response time of \textbf{1.6920 ms}. This is a critical threshold for industrial control loops, ensuring that malicious signatures are extracted and mitigated within the \textbf{10 ms real-time window} required by standard Industrial Control Systems (ICS).

\subsubsection{Hardware-Aware Structural Efficiency}
A key barrier to FPGA and ASIC deployment in IIoT is the "memory wall"—the latency and power penalty incurred by external DDR memory access. BGA bypasses this bottleneck through its lightweight design (\textbf{123.4 K parameters}). Requiring only \textbf{0.471 MB} of storage, the entire model can be mapped onto the on-chip \textbf{Block RAM (BRAM)} of mid-range FPGAs. This structural efficiency ensures deterministic latency, which is paramount for mission-critical infrastructures like gas pipelines where stochastic processing delays can lead to synchronization failures and physical-cyber safety misalignment.BGA provides a reliable defense for real-time industrial edge operations by guaranteeing a sub-10ms response on ARM processors. This efficiency ensures that malicious signatures are mitigated within the strict windows required by Industrial Control Systems (ICS), which is essential for operational safety.

\subsubsection{Substantive Significance of High-Precision Improvements}
To bridge the gap between numerical metrics and operational reality, we evaluate the substantive impact of BGA's precision gains. In a typical high-throughput industrial mesh processing 1,000,000 flow records per hour, the 0.6\% F1-score improvement over standard LSTM baselines translates to \textbf{6,000 fewer security errors per hour}. Given that malicious activities in IIoT, such as MSCI, are often stealthy and infrequent, this improvement ensures the detection of long-tail attack patterns that would otherwise evade defense. 

The combination of this high fidelity with a low parameter count and sub-10ms response times indicates that BGA possesses Pareto-optimal characteristics relative to the tested recurrent and attention-based baselines. As demonstrated in our hardware simulation (Section 4.3.7), while the latency increases on low-power platforms, BGA maintains a response time of 1.6920 ms, which remains safely below the 10 ms real-time threshold required by Industrial Control Systems (ICS). For critical infrastructure, where the cost of a single missed command can result in physical destruction, every 0.1\% increase in detection fidelity, supported by guaranteed real-time edge feasibility, directly enhances the system's \textbf{Safety Integrity Level (SIL)} and effectively bridges the gap between deep sequence modeling and industrial edge defense.

\subsection{Comparative Analysis with Evaluated Transformer-based Architectures}
While Transformer-based language models have revolutionized NLP, our quantitative benchmarking (see Section \ref{Experiments and Analysis} and Table \ref{tab:llm_benchmark}) confirms that their application to real-time traffic detection at the IIoT edge remains impractical under the evaluated configurations for several fundamental reasons:

\begin{itemize}
    \item \textbf{Computational Disparity:} Even heavily quantized models like the evaluated TinyBERT baseline require over 14 million parameters. Our experiments show that BGA is 110$\times$ more memory-efficient, making it suitable for low-power ARM-based gateways that lack the VRAM required for standard Transformer weights.
    \item \textbf{Throughput Bottleneck:} BGA’s 0.2820 ms latency is \textbf{14.9 times faster} than the tested TinyBERT configuration. In high-speed industrial networks processing millions of packets per second, the serial nature of self-attention in such Transformer-based models leads to massive packet drops, whereas BGA satisfies the most stringent microsecond-level real-time requirements.
    \item \textbf{Hardware-Contextualized Latency Scaling:} A critical distinction must be made between research-grade environments and operational hardware constraints. While TinyBERT reports a 4.1946~ms latency on a high-performance PC, this figure is deceptive for edge deployment. As demonstrated in our 1.2~GHz ARM simulation (Section 4.3.7), BGA maintains a viable 1.6920~ms response time under extreme constraints. In contrast, scaling the evaluated Transformer’s performance to the same platform would result in latencies exceeding 25~ms, far surpassing the 10~ms industrial safety threshold.
    \item \textbf{Feature Precision:} Transformer-based models are often pre-trained on semantic text. Conversely, IIoT traffic consists of high-precision numerical sequences. The evaluated generic Transformer baselines often lack the sensitivity to detect the 0.1\% deviations in industrial setpoints that signify a stealthy attack, frequently treating such critical process anomalies as linguistic noise.
\end{itemize}

In summary, these findings serve as a representative case study highlighting the suitability of BGA compared to specific compressed Transformer configurations, rather than claiming definitive superiority over all modern sequence modeling paradigms.

\subsection{Comparison with Emerging Sequence Models (2024-2026)}
While recent breakthroughs in sequence modeling, such as Mamba-2 \citep{ref33} (Generalized State Space Models), RWKV-v6 \citep{ref34} (Linear Attention), and emerging pre-trained traffic transformers \citep{ref35}, have redefined the performance ceilings in the 2024-2026 research landscape, their direct application to IIoT encrypted traffic detection faces several practical hurdles. 

First, the evolved Mamba-2 and RWKV-v6 architectures are optimized for processing extremely long context windows in NLP; however, encrypted network flows are typically represented by relatively short, high-entropy feature sequences where long-range dependencies are less critical than immediate noise suppression. Second, the deployment of these recent architectures often relies on specialized software kernels (e.g., customized CUDA operators for Mamba-2), which are largely incompatible with the heterogeneous, low-power embedded systems found in industrial edge gateways. Lastly, the BGA framework prioritizes 'Neural Distillation' through gated residuals, which is more effective at decoupling deterministic attack signatures from the stochastic jitter of TLS 1.3 encryption than general-purpose sequence models. By maintaining a microsecond-level latency (0.2820 ms) and a lightweight parameter count (130.6K), BGA provides a more a favorable balance solution for real-time edge resilience in the 2025-2026 deployment environment than these high-complexity mainstream architectures.

While emerging architectures such as Mamba-2 and RWKV-v6 offer promising efficiency for long-context sequences, their specific performance in the high-entropy, short-sequence IIoT traffic domain remains a subject for future empirical investigation. Therefore, the efficiency advantages of BGA identified in this study should be interpreted within the scope of the currently tested baselines.

\subsection{Ecological Validity and Resilience to Protocol Obfuscation}
While the Gaussian noise stress tests in Section 4.4.3 demonstrate BGA's structural resilience to statistical uncertainty, the framework's \textit{ecological validity} is further grounded in its capacity to handle protocol-level obfuscation. In modern encrypted flows like TLS 1.3, random padding and dummy packets are utilized to obscure packet lengths and timing signatures. However, these mechanisms primarily inject stochastic entropy into the feature sequence. 

The BGA framework mitigates these artifacts through its Neural Distillation logic. Specifically, the adaptive gating mechanism assigns low importance coefficients to feature channels dominated by randomized padding, while the BiLSTM layer preserves the underlying deterministic "behavioral rhythm" of the industrial control commands. By bridging this visibility gap, BGA provides a robust mechanism for maintaining physical process integrity in the face of sophisticated, hidden adversarial behaviors.

Furthermore, the BGA framework is architected for graceful degradation in degraded data environments. Should primary industrial control features, such as \textit{setpoints}, become entirely unavailable, the ANOVA pre-filter automatically re-calibrates to identify the next highest-discriminatory behavioral signatures, such as inter-arrival time (IAT) statistics or packet burst rhythms. This structural flexibility, enabled by the synergy between automated feature ranking and neural distillation, allows the system to maintain a robust defensive posture even under limited visibility or partial feature observation.

\section{Conclusion} \label{Conclusion}

\subsection{Summary of Contributions}
In this work, we proposed BGA, a noise-immune neural distillation framework for encrypted industrial threat intelligence. By integrating WGAN-GP for manifold reconstruction and an adaptive gated BiLSTM-Attention architecture, we successfully addressed the challenges of class imbalance and "attention dilution" in high-entropy TLS 1.3 flows. Experimental results on CIC-IDS-2018 and Edge-IIoT benchmarks confirmed that BGA achieves a performance ceiling of over 95\% accuracy with an ultra-low inference latency of 0.2820 ms, providing a robust and real-time defense mechanism for industrial edge gateways.

\subsection{Limitations and Deployment Risks}
Despite its superior performance, the BGA framework possesses inherent limitations. First, the fidelity of WGAN-GP augmentation is contingent upon the availability of high-quality "seed" samples; if the initial minority class data is too sparse to represent the true adversarial manifold, the generated samples may introduce systemic bias. Second, our analysis of failure cases indicates that BGA may struggle with "stealthy logical drift" attacks, where adversaries inject commands that reside within legitimate setpoint boundaries but violate higher-level process interdependencies. 

Furthermore, a significant constraint regarding our architectural assessment is the reliance on theoretical scaling and simulation for assessing hardware feasibility. Although the low parameter count and estimated latency are promising, the absence of physical end-to-end testing on actual industrial ARM gateways or embedded devices means that these results should be interpreted as architectural potentials rather than definitive deployment evidence. 

From a \textbf{deployment perspective}, the primary risk involves the computational overhead on extremely low-end embedded devices when managing massive concurrent flows. While our 1.6920 ms simulated latency is viable for standard gateways, a sudden burst of high-density traffic could lead to buffer overflows or synchronization jitters in time-sensitive 5G meshes.

\subsection{Future Research Directions}
To further advance the field of post-payload threat intelligence, we propose several forward-looking research directions:
\begin{enumerate}
    \item \textbf{Physical Deployment and Hardware-in-the-Loop Validation:} The foremost objective of our future work is to transition from theoretical simulation to empirical validation on actual industrial gateways and ARM-based embedded devices. This will involve conducting end-to-end testing to analyze the performance impact of OS-level task scheduling, real-time interrupt handling, and memory bus contention, thereby confirming the BGA framework's operational resilience in physical industrial environments.
    \item \textbf{Relational Spatio-Temporal Modeling:} Future work should explore the integration of \textbf{Graph Neural Networks (GNNs)} with BGA's gated distillation. This would enable the model to capture not only individual flow signatures but also the complex relational dependencies between multiple network entities.
    \item \textbf{Privacy-Preserving Threat Intelligence:} Implementing \textbf{Federated Learning} paradigms could allow heterogeneous industrial nodes to collaboratively train BGA models without exposing sensitive raw traffic data, addressing the critical trade-off between security and data privacy.
    \item \textbf{6G and Quantum-Resilient Security:} As industrial networks evolve toward 6G, research should focus on optimizing neural distillation for ultra-massive machine-type communications and investigating BGA's resilience against quantum-based protocol obfuscation.
\end{enumerate}

By addressing these challenges, the research community can bridge the remaining visibility gaps in ubiquitous encryption ecosystems, ensuring the long-term integrity of critical physical infrastructures.

\section*{Acknowledgements}
The authors acknowledge the National Key Research and Development Program of China (Grant No. 2022YFB3103602) for the financial support.

\section*{Declaration of Competing Interest}
The authors declare that they have no known competing financial interests or personal relationships that could have appeared to influence the work reported in this paper.

\section*{Author Contributions (CRediT)}
\textbf{Sheng Hong}: Conceptualization, Methodology, Funding acquisition, Supervision. 
\textbf{Yixuan Huang}: Software, Data curation, Validation, Writing - original draft. 
\textbf{Weiwei Jiang}: Methodology, Formal analysis, Writing - review \& editing.
\textbf{Junyuan Zhang}: Investigation, Methodology, Validation. 
\textbf{Jiacheng Wang}: Resources, Software, Validation. 
\textbf{Ruijian Jiao}: Formal analysis, Visualization, Writing - review \& editing.

\section*{Data Availability}
The data supporting the findings of this study are proprietary and confidential. Due to intellectual property restrictions and non-disclosure agreements, the authors do not have the permission to share the underlying raw data.

\bibliographystyle{cas-model2-names} 
\bibliography{main}


\bio{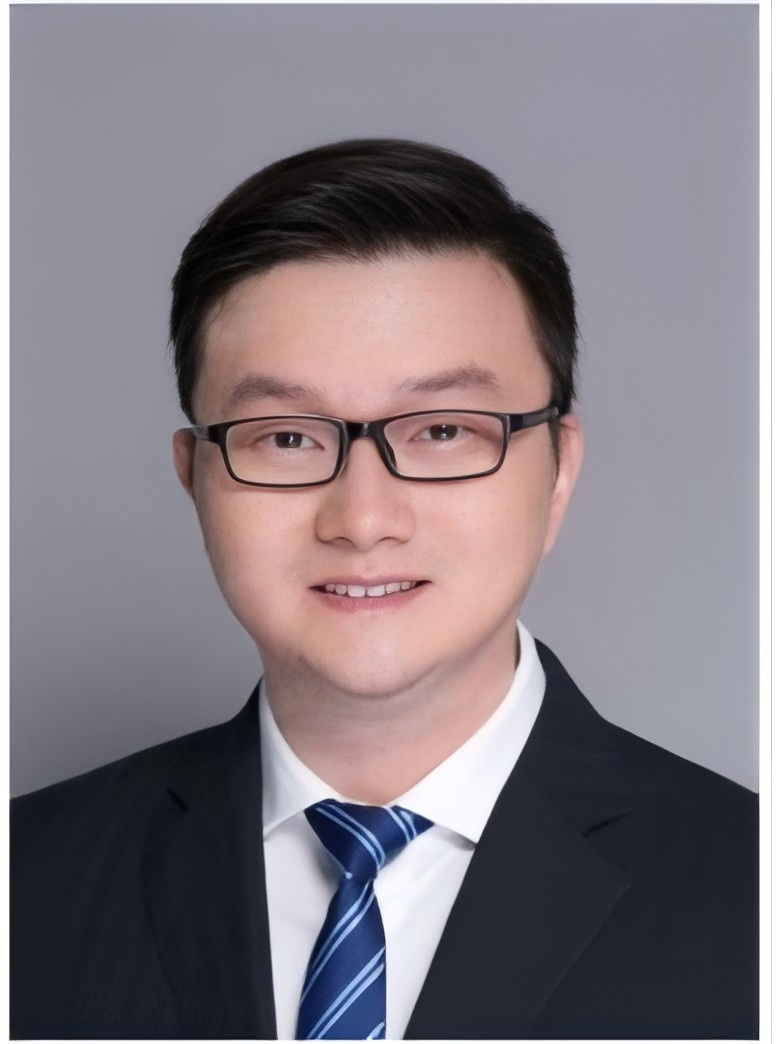}
{\bf Sheng Hong} received his Ph.D. from Beihang University and expanded his research horizon as a Visiting Scientist at the Georgia Institute of Technology, USA. He is currently an Associate Professor and Doctoral Supervisor at Beihang University. As a distinguished Beijing Subject Leader and recipient of the First Prize in Science and Technology Progress, Dr. Hong has a proven track record of leadership in the scientific community. As a Principal Investigator, he has spearheaded numerous prestigious initiatives, including the National Key R\&D Program of China and multiple grants from the National Natural Science Foundation of China (NSFC). His research interests include artificial intelligence and big data, AI-driven cyber security and industrial internet.\\
E-mail: shenghong@buaa.edu.cn
\endbio

\vspace{25pt}

\bio{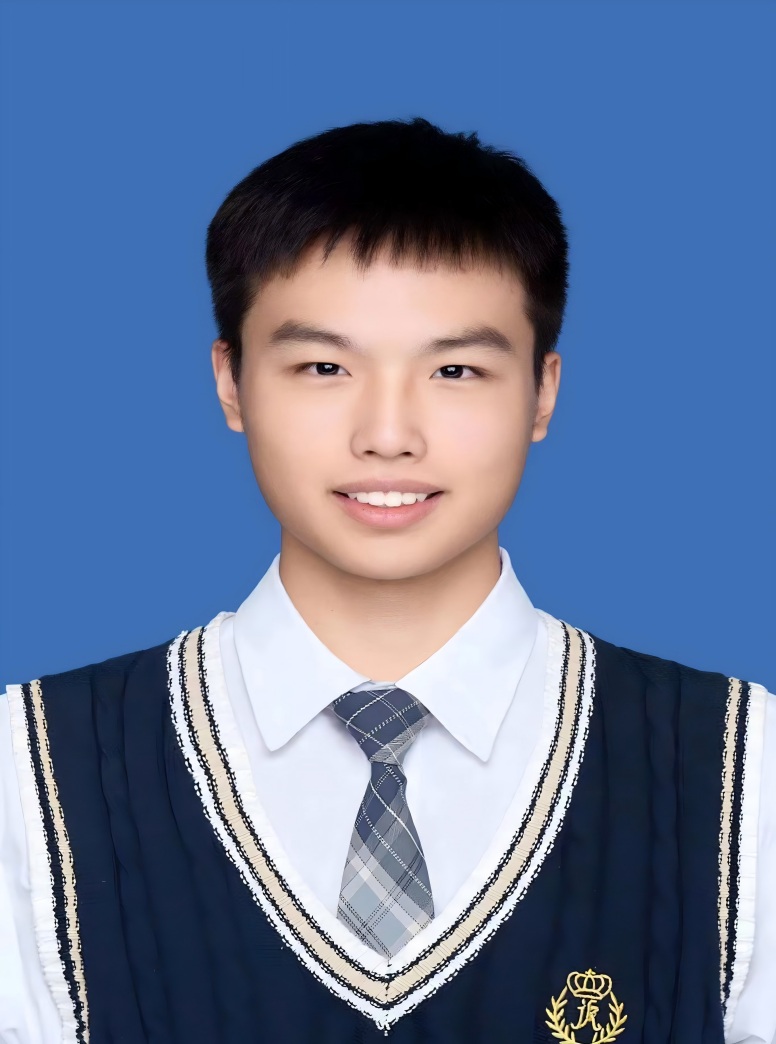}
{\bf HUANG Yixuan} was born in 2005. He is currently pursuing his B.S. degree in Cyber Science and Technology at Beihang University. His research interests include network intrusion detection, adversarial ML, and the implementation of post-quantum cryptosystems.\\
E-mail: 23371296@buaa.edu.cn
\endbio

\vspace{25pt}

\bio{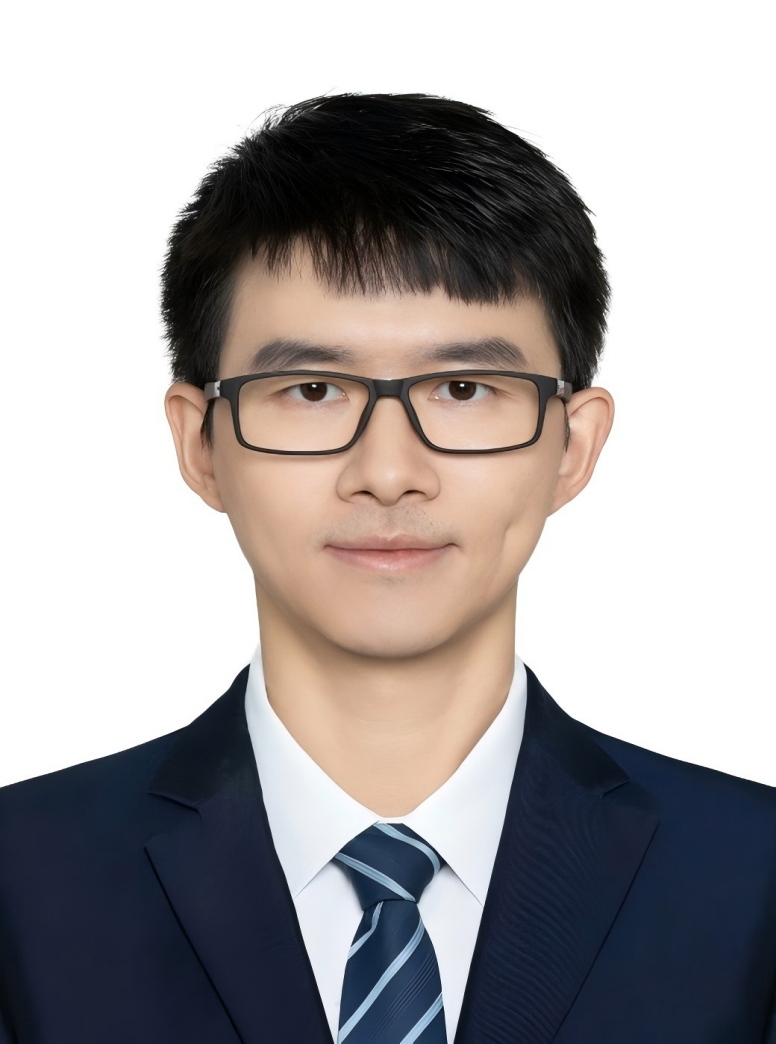}
{\bf Weiwei Jiang} (IEEE Senior Member) received the B.Sc. Degree of Electronic Engineering and Ph.D. Degree of Information and Communication Engineering from the Department of Electronic Engineering, Tsinghua University, Beijing, China, in 2013 and 2018, respectively. He is currently an associate professor with the School of Information and Communication Engineering, Beijing University of Posts and Telecommunications, and Key Laboratory of Universal Wireless Communications, Ministry of Education. His current research interests include artificial intelligence, machine learning, big data, wireless communication and edge computing. He has published more than 100 academic papers in IEEE Trans and other journals, with more than 5900 citations in Google Scholar. He is one of 2022, 2023, 2024 and 2025 Stanford's List of World's Top 2\% Scientists.\\
E-mail: jww@bupt.edu.cn
\endbio

\vspace{25pt}

\bio{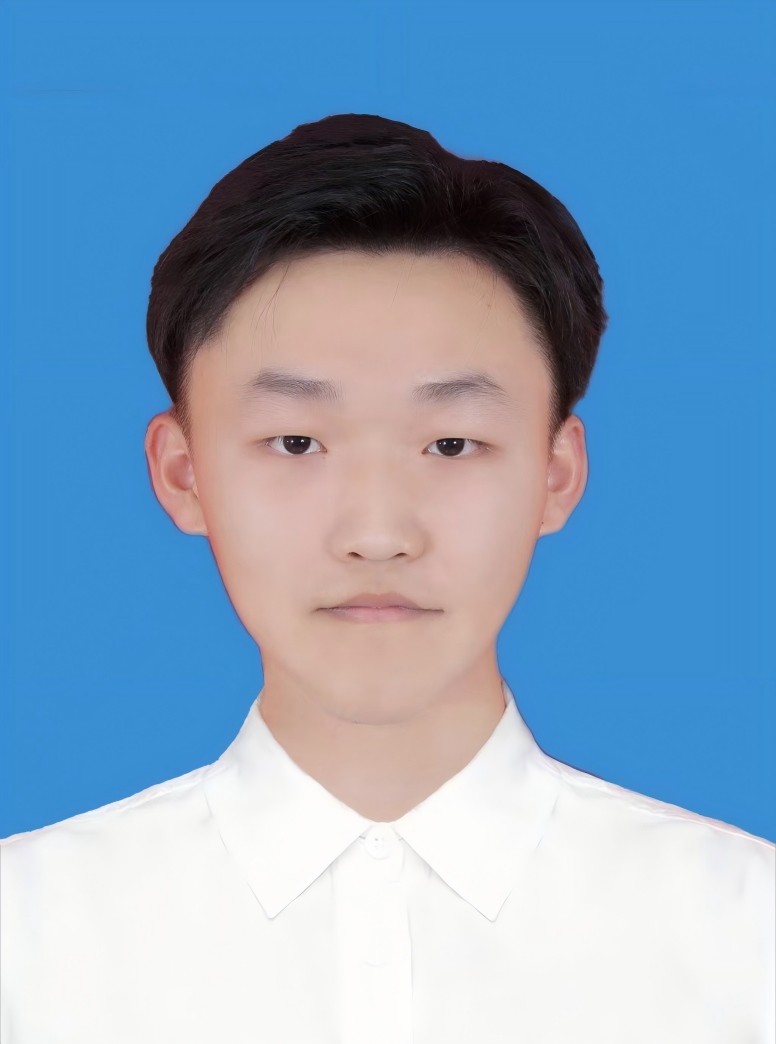}
{\bf ZHANG Junyuan} was born in 2001. He received his bachelor's degree in engineering from Inner Mongolia University of Technology, China, in 2025. He is currently pursuing his master's degree at Beijing Electronic Science and Technology Institute. His research interests include artificial intelligence and big data, quantum key distribution and post-quantum cryptography.\\
E-mail: 1943351829@qq.com
\endbio

\vspace{25pt}

\bio{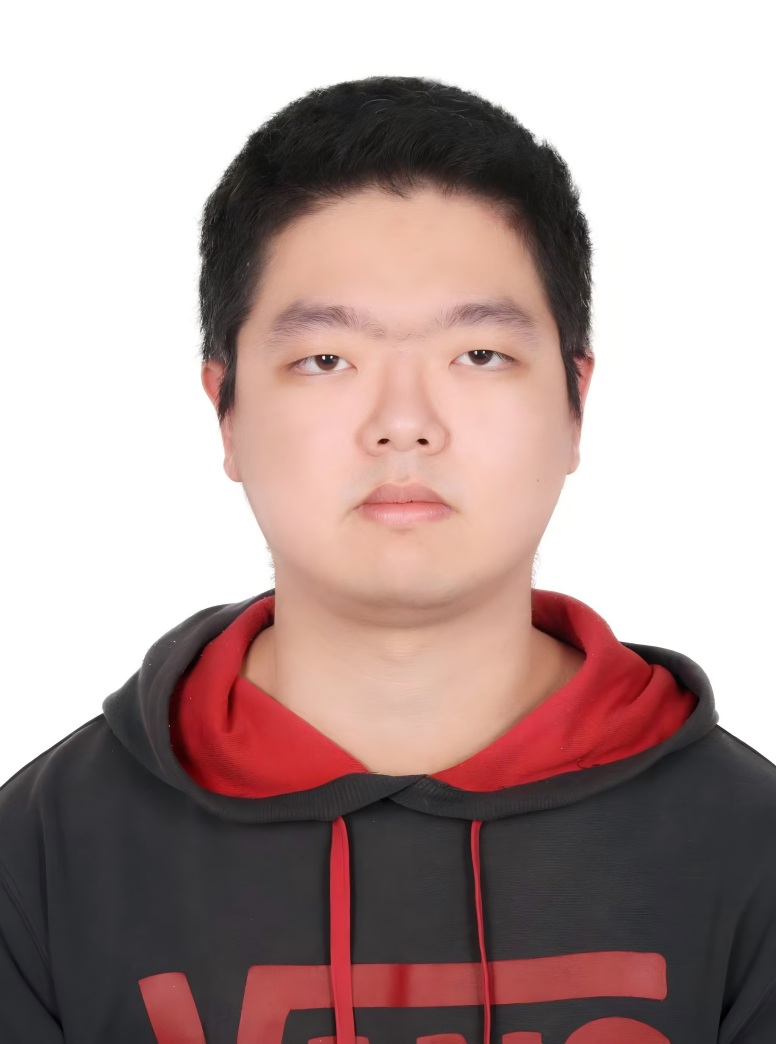}
{\bf WANG Jiacheng} was born in 2001. He received his B.S. degree in information security from Lanzhou University, China, in 2023. He is currently pursuing his M.Eng degree at the School of Cyber Science and Technology, Beihang University, China. His research interests include artificial intelligence, information security, and applied cryptography.\\
E-mail: wjc1321@163.com
\endbio

\vspace{25pt}

\bio{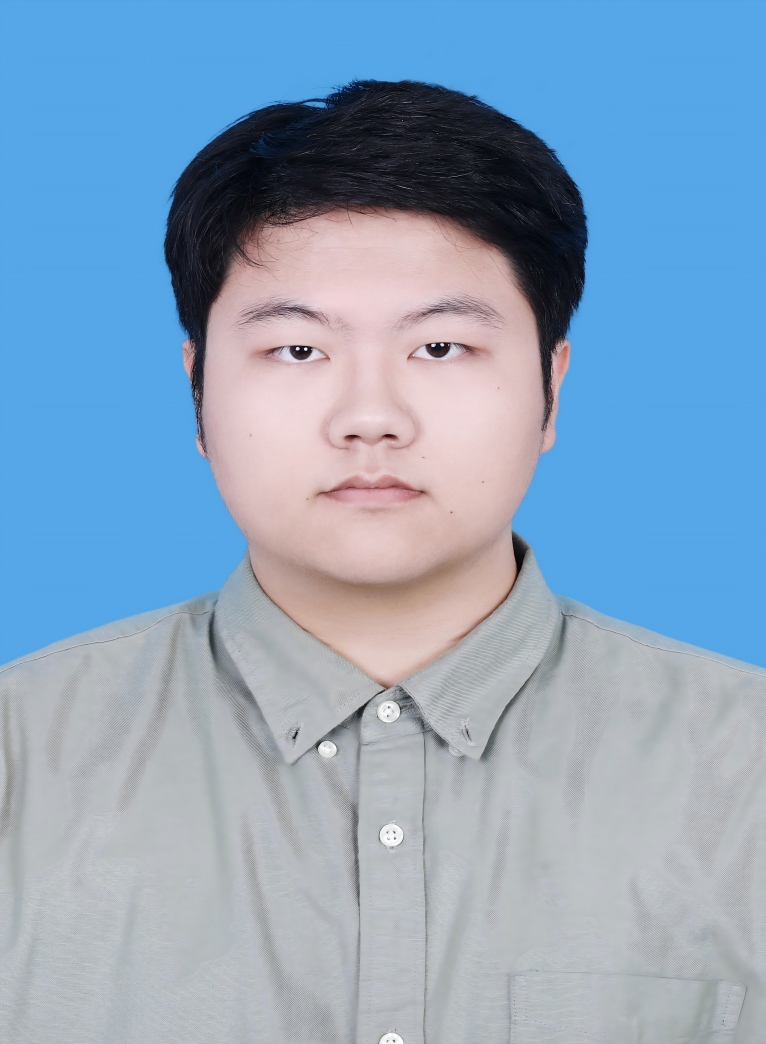}
{\bf JIAO Ruijian} was born in 2003. He received his B.S. degree from Beihang University, China, in 2021. He is currently pursuing his M.S. degree at Beihang University. His research interests include deep learning for cybersecurity, encrypted traffic and intelligent detection systems.\\
E-mail: w\_nter\_@outlook.com
\endbio

\end{document}